\documentclass[twocolumn]{aastex631}

\shorttitle{LG monitoring. V}
\shortauthors{Parto et al.}

\usepackage{float}
\usepackage[utf8]{inputenc}
\DeclareUnicodeCharacter{2212}{\textendash}
\begin{document}

\title{THE ISAAC NEWTON TELESCOPE MONITORING SURVEY OF LOCAL GROUP DWARF GALAXIES--V.
THE STAR FORMATION HISTORY OF SAGITTARIUS DWARF IRREGULAR GALAXY DERIVED FROM LONG PERIOD VARIABLE STARS}

\author{Tahere Parto}
\affiliation{School of Astronomy, Institute for Research in Fundamental Sciences (IPM), P.O. Box 1956836613, Tehran, Iran}
\affiliation{Physics Department, Alzahra University, Vanak, 1993891176, Tehran, Iran}

\author{Shahrzad Dehghani}
\affiliation{Department of Physics, University of Cologne, Cologne, Germany}

\author{Atefeh Javadi}
\affiliation{School of Astronomy, Institute for Research in Fundamental Sciences (IPM), P.O. Box 1956836613, Tehran, Iran}

\author{Elham Saremi}
\affiliation{School of Astronomy, Institute for Research in Fundamental Sciences (IPM), P.O. Box 1956836613, Tehran, Iran}
\affiliation{Instituto de Astrofísica de Canarias, Vía Láctea s/n, 38205 La Laguna, Tenerife, Spain}
\affiliation{Departamento de Astrofísica, Universidad de La Laguna, 38205 La Laguna, Tenerife, Spain}

\author{Jacco Th. van Loon}
\affiliation{Lennard-Jones Laboratories, Keele University, ST5 5BG, UK}

\author{Habib G. Khosroshahi}
\affiliation{School of Astronomy, Institute for Research in Fundamental Sciences (IPM), P.O. Box 1956836613, Tehran, Iran}
\affiliation{Iranian National Observatory, Institute for Research in
	Fundamental Sciences (IPM), Tehran, Iran}

\author{Iain McDonald}
\affiliation{Jodrell Bank Centre for Astrophysics, Alan Turing Building, University of Manchester, M13 9PL, UK}

\author{Mohammad T. Mirtorabi}
\affiliation{Physics Department, Alzahra University, Vanak, 1993891176, Tehran, Iran}

\author{Mahdieh Navabi}
\affiliation{School of Astronomy, Institute for Research in Funda-
	mental Sciences (IPM), P.O. Box 1956836613, Tehran, Iran}

\author{Maryam Saberi}
\affiliation{Rosseland Centre for Solar Physics, University of Oslo, P.O. Box 1029 Blindern, NO-0315 Oslo, Norway}
\affiliation{Institute of Theoretical Astrophysics, University of Oslo, P.O. Box 1029 Blindern, NO-0315 Oslo, Norway}


\email{atefeh@ipm.ir}

\begin{abstract}
We conducted an optical monitoring survey of the Sagittarius dwarf irregular galaxy (SagDIG) during the period of June 2016 -- October 2017, using the 2.5-m Isaac Newton Telescope (INT) at La Palama. Our goal was to identify Long Period Variable stars (LPVs), namely asymptotic giant branch stars (AGBs) and red supergiant stars (RSGs), to obtain the Star Formation History (SFH) of isolated, metal-poor SagDIG. For our purpose, we used a method that relies on evaluating the relation between luminosity and the birth mass of these most evolved stars.
We found $27$ LPV candidates within two half-light radii of SagDIG. $10$ LPV candidates were in common with previous studies, including one very dusty AGB (x-AGB).
By adopting the metallicity $Z = 0.0002$ for older population and $Z=0.0004$ for younger ages, we estimated that the star formation rate changes from $0.0005\pm0.0002$ M$_{\odot}$yr$^{-1}$kpc$^{-2}$ ($13$ Gyr ago) to $0.0021 \pm 0.0010$ M$_{\odot}$yr$^{-1}$kpc$^{-2}$ ($0.06$ Gyr ago).
Like many dwarf irregular galaxies, SagDIG has had continuous star formation activity across its lifetime, though with different rates, and experiences an enhancement of star formation since $z \simeq 1$. We also evaluated the total stellar mass within two half-light radii of SagDIG for three choices of metallicities. For metallicity $Z = 0.0002$ and $Z=0.0004$ we estimated the stellar mass M$_*$ = ($5.4 \pm 2.3$) $\times$ $10^ 6$  and ($3.0 \pm 1.3$) $\times$ $10^ 6$ M$_{\odot}$, respectively. Additionally, we determined a distance modulus $\mu$ = $25.27\pm0.05$ mag, using the tip of the red giant branch (TRGB).

\end{abstract}


\section{Introduction}
\label{introduction}

The physical properties of dwarf galaxies make them excellent objects for studying the formation and evolution of galaxies.
They are the most dark matter dominated galaxies and the simplest systems that demonstrate how dark matter works on small scales.
Furthermore, they are located at extreme limits for the formation of galaxies in terms of size, mass, and metallicity \cite[]{2019ARA&A..57..375S}.
The internal and environmental mechanisms can severely affect these low mass systems, which can be studied to determine the role of different processes and their efficiencies in the evolution of galaxies \cite[]{2019IAUS..344....3H}.
Among different morphologies of dwarf galaxies, dwarf irregulars (dIrrs) are typically gas-rich. They can be used for investigating the relation between stellar and gas-phase metallicities to see how gas flows shape the metallicity distribution and to what extent the high gas fraction can affect their metallicities \cite[]{2017ApJ...834....9K}.
They are also a key to answering questions such as the efficiency of star formation in low gas densities, the importance of sequential star formation, and the possible role of star formation in creating breaks in the structure of the outer regions of spiral galaxies \cite[]{2012AJ....144..134H}.
Low metallicity dIrrs resemble the building blocks of the Universe and reveal the efficiency of different astrophysical processes in forming the first stars.
Moreover, they help figure out the questions about dust formation in the early Universe \citep[e.g.,][]{Boyer_2014}.

The Local Group (LG) offers advantages to the study of dwarf galaxies due to their proximity and their range in mass, morphology, and age. Among LG dwarfs, the Sagittarius dwarf irregular galaxy (SagDIG) has unique features that make understanding its formation history a topic of high interest. The first and foremost feature is its low metallicity $Z=0.00025$ \cite[]{2002A&A...384..393M}. Combined with its close distance $1.07 \pm 0.09$ Mpc \cite[]{2002A&A...384..393M}, it allows finding resolved metal-deficient stars. The other particular property of SagDIG is its notable gas fraction, $M($H\,{\sc i}$)/M_*=4.6\pm1.2$ \cite[]{2017ApJ...834....9K}. It leads to a high star formation rate (SFR), making SagDIG the most rapidly growing galaxy among LG dwarf galaxies \cite[]{2017ApJ...834....9K}. 
Moreover, the isolation of SagDIG makes it an ideal laboratory to study how a low mass galaxy would evolve in the absence of environmental effects \cite[]{2014ApJ...789..147W}.

SagDIG was found by \cite{1978A&A....65..153C} and \cite{1978MNRAS.183P..97L} with the 1 m ESO Schmidt telescope and through the UK Schmidt IIIaJ Southern Sky Survey, respectively. Later inspections of this object revealed its irregular shape \citep[]{1978A&A....65..153C,1978MNRAS.183P..97L}. Various studies calculate its distance modulus using the tip of red giant branch stars (TRGB). To name a few, \citet{2002A&A...384..393M} found the distance modulus $\mu$ = $25.14\pm0.18$ mag, \citet{2014ApJ...789..147W} estimated  $\mu$ = $25.11$ mag, and \citet{2016MNRAS.458.1678H} estmated $\mu$ = $25.36\pm0.15$ mag.
In this study we adopted $\mu$ = $25.27\pm0.05$ mag, which we estimated using the same method in Section \ref{trgb}.

The star formation history (SFH) of SagDIG has been estimated in many studies. \citet{1999A&A...352..363K} estimated the SFH of SagDIG through classifying and counting stars in the young and old populations based on their color and position on the [$I$, $V$--$I$] color--magnitude diagram (CMD) and employing a synthetic CMD. They found that SagDIG has an ongoing star formation similar to other dwarf irregulars such as NGC 6822, Pegasus, Sextans A, and Antlia with a rate ten times higher than average of its whole lifetime SFR.

\citet{2005A&A...439..111M} used the HST/ACS deep images and identified a well-populated RGB. They concluded that the main population of this galaxy are stars older than $1$ Gyr. They also found populations of main-sequence stars, He-burning blue loop stars, blue and red supergiants, and a population of AGBs that show extended star formation in this galaxy. Moreover, using the luminosity function of main sequence stars, they estimated an SFR for the last few hundred mega years and concluded that the SFR is mainly constant in this time interval.
\citet{2014ApJ...789..147W} used the HST archival images to study the SFHs of 40 Local Group galaxies, including SagDIG. They constructed a model CMD for each galaxy and matched it to the observed CMD to derive the cumulative SFH with random and systematic errors. We
will compare our results with these studies in Section \ref{sec:science}.

It is also possible to investigate the history of a galaxy with long-period variable stars (LPVs), namely asymptotic giant branch stars (AGBs) and massive red supergiants (RSGs). LPVs are the most evolved stars with strong radial pulsation due to variation in the opacity and radiation transport, leading to a fluctuation in their brightness within $300$ -- $1200$ days. Although different types of variable stars could be beneficial for the aim of the SFH survey, the use of LPVs is more advantageous. Firstly, LPVs are at the last stage of evolution, and the relation between stars' luminosity and birth-mass can be applied to estimate LPV's mass. Secondly, LPVs cover a diverse range of ages between $30$ Myr to $10$ Gyr old. Thirdly, LPVs' brightness in near-infrared is less affected by circumstellar extinction \citep{2011MNRAS.411..263J, 8214023}. 

This work is part of the Isaac Newton Telescope monitoring survey of Local Group dwarf galaxies.
Various types of dwarf galaxies, including 43 dwarf spheroidals, six dwarf irregulars, six dwarf transition galaxies and four globular clusters come under the scrutiny of our team \citep{2017JPhCS.869a2068S, 2021arXiv210110874P}. The main objectives of this survey include: identifying all LPVs in the dwarf galaxies of the LG observable in the northern hemisphere; obtaining accurate time-averaged photometry for all LPVs; constructing the SFHs from LPVs luminosity distribution; obtaining their pulsation amplitude; modeling their SEDs; and studying their mass loss as a function of stellar properties like mass, luminosity, metallicity, and pulsation amplitude \cite[]{2020ApJ...894..135S}. This paper (paper V) focuses on constructing the SFH of SagDIG.
Paper I \cite[]{2020ApJ...894..135S} introduced the survey and the detection of LPVs in Andromeda I (And I), Paper II \cite[]{Saremi_2021} presents the results of SFH in And I. Paper III (in preparation) discusses the role of LPVs in the chemical enrichment of And I, and paper IV \cite[]{Navabi_2021} presents  SFH in Andromeda VII (And VII).

This paper is organized as follows: In Section \ref{sec:observations} we describe observations, data reduction, photometry and contamination. Besides, we compare our catalog with previous catalogs of SagDIG. In Section \ref{sec:var} we present the method to find LPVs. In Section \ref{trgb} we estimate the distance modulus using TRGB method. In Section \ref{sec:Star Formation History} we introduce the method we use for constructing the SFH. In Section \ref{sec:science} we discuss our results, followed by a summary and conclusion in Section \ref{sec:conclusions}.

\begin{figure*}[t]
	\makebox[\textwidth][c]{\includegraphics[width=1.2\textwidth]{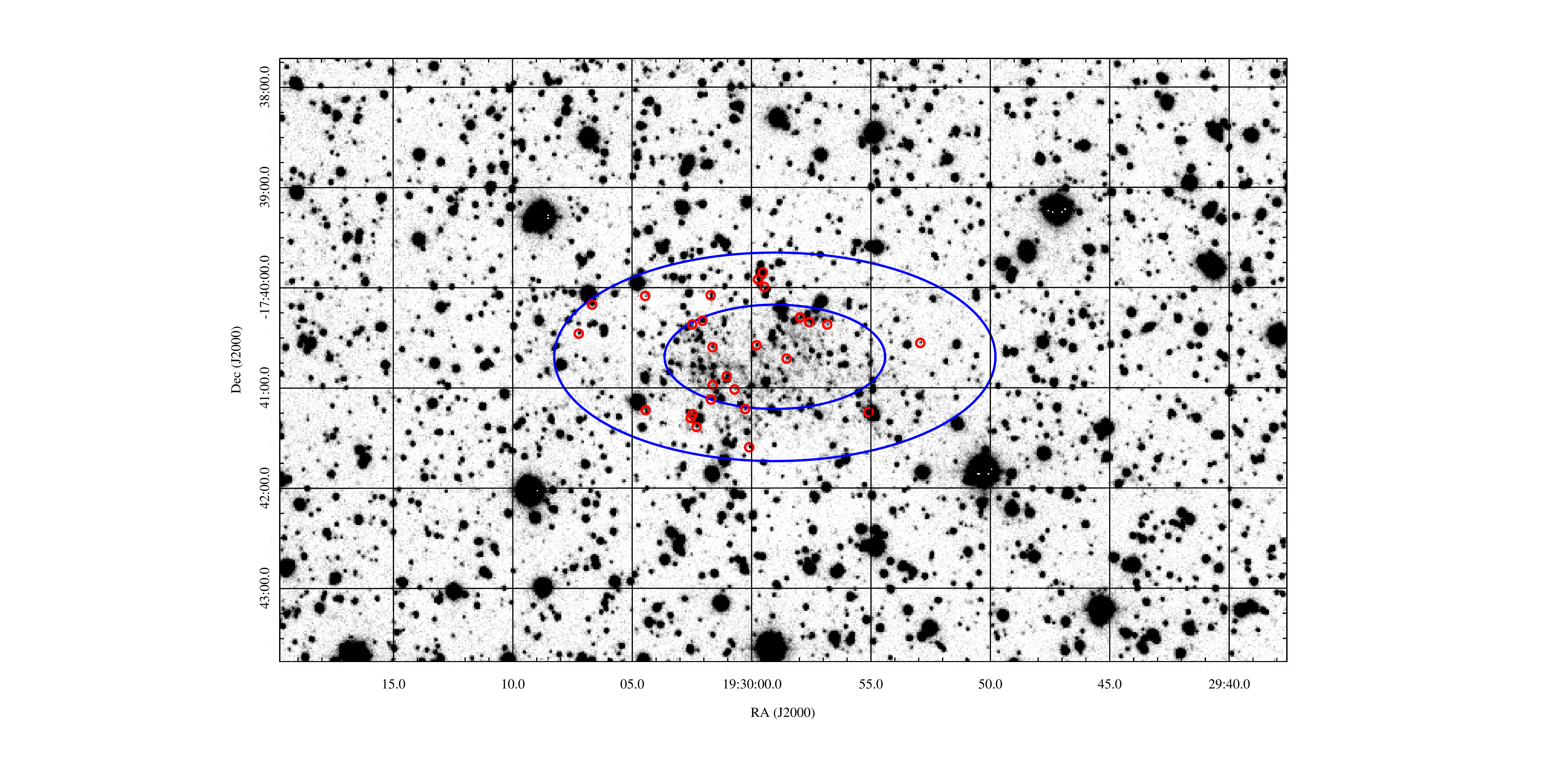}}%
	\caption{The master WFC image of SagDIG galaxy restricted to a field of $10\rlap{}^\prime \times 6\rlap{}^\prime$. The LPV candidates are marked with red circles.
		Blue ellipses show the half and two half-light radii of SagDIG \cite[]{2014A&A...570A..78B} with the ellipticity $0.50$ and position angle $90^\circ$ \cite[]{2012AJ....144....4M}.  }
	\label{fig:SagDIG}
\end{figure*}

\section{Observation and Data processing}
\label{sec:observations}
\begin{table}
	\centering
	\caption{\label{table:tdata} Log of WFC observations of SagDIG}
	\label{tab:log}
	\begin{tabular}{cccccc}
		\hline
		\hline
		Date    & Epoch & Filter & $t_{\rm exp}$ & seeing & Airmass \\
		(y m d) &       &        & (sec)       & arcsec & \\
		\hline
		\hline
		2016 06 13 & 1 & i &  719 & 2.61 & 1.711 \\
		2016 08 10 & 2 & i &  555 & 1.23 & 1.564 \\
		2016 08 10 & 2 & V &  735 & 1.64 & 1.471 \\
		2016 08 12 & 2 & i &  558 & 1.37 & 1.449 \\
		2016 10 21 & 3 & i & 2715 & 1.84 & 1.595 \\
		2017 08 02 & 4 & i &  555 & 1.44 & 1.453 \\
		2017 08 02 & 4 & V &  735 & 1.37 & 1.450 \\
		2017 09 02 & 5 & i &  555 & 1.08 & 1.457 \\
		2017 09 02 & 5 & V &  735 & 1.18 & 1.449 \\
		2017 10 06 & 6 & i &  555 & 1.61 & 1.511 \\
		2017 10 08 & 6 & V &  735 & 1.82 & 1.542 \\
		
		\hline
		\hline
	\end{tabular}
	\vspace{1ex}

\end{table}

Images of SagDIG in the $i$ and $V$-band were obtained from June $2016$ to October $2017$ using the Isaac Newton Telescope (INT) Wide Field Camera (WFC) that includes $4$ CCDs with $4100 \times 2046$ pixel dimension and
a pixel-scale of $0.33$ arcsec/pixel. The observing log is presented in Table \ref{table:tdata}. In Fig.~\ref{fig:SagDIG}, the master image in $i$ and $V$-band, restricted to a field of $10\rlap{}^\prime \times 6\rlap{}^\prime$, is presented. The coordinate center of SagDIG is located in the center of the central CCD (CCD 4). Since more than twice the half-light radii of SagDIG is covered in CCD 4, we only considered this CCD for the photometry procedure.

The WFC images were reduced by the THELI image processing pipeline \cite[]{2005AN....326..432E}. We performed the photometry process in both filters using the DAOPHOT package \cite[]{1987PASP...99..191S}. For this purpose, we used the {\sc DAOPHOT} routine to select $40$ isolated stars in different positions on the field and built a constant point spread function (PSF) model for each image. A master image was made by combining single images (using {\sc DAOMATCH}, {\sc DAOMASTER}, and {\sc MONTAGE2} routines) to generate a star list (using the {\sc ALLSTAR} routine). The {\sc ALLFRAME} routine uses the star list to estimate the instrumental magnitudes of stars by fitting the PSF models in individual images \cite[]{1994PASP..106..250S}.

The transformation of the instrumental magnitudes onto the standard system was carried out using observations of standard stars \cite[]{1992AJ....104..340L} and the {\sc NEWTRIAL}  routine \cite[]{1996PASP..108..851S}. The final catalog consists of $12\,538$ stars in the field of CCD 4 and $678$ stars in the $2r_{\rm h}$ from the center of SagDIG, assuming the half-light radius $r_{\rm h} =1\rlap{.}^\prime1$ \cite[]{2014A&A...570A..78B}. Additional details on the observations and photometry procedures are provided in \citet{2020ApJ...894..135S}.

\begin{figure}[t]
	\centering
	\includegraphics[width=\linewidth, clip]{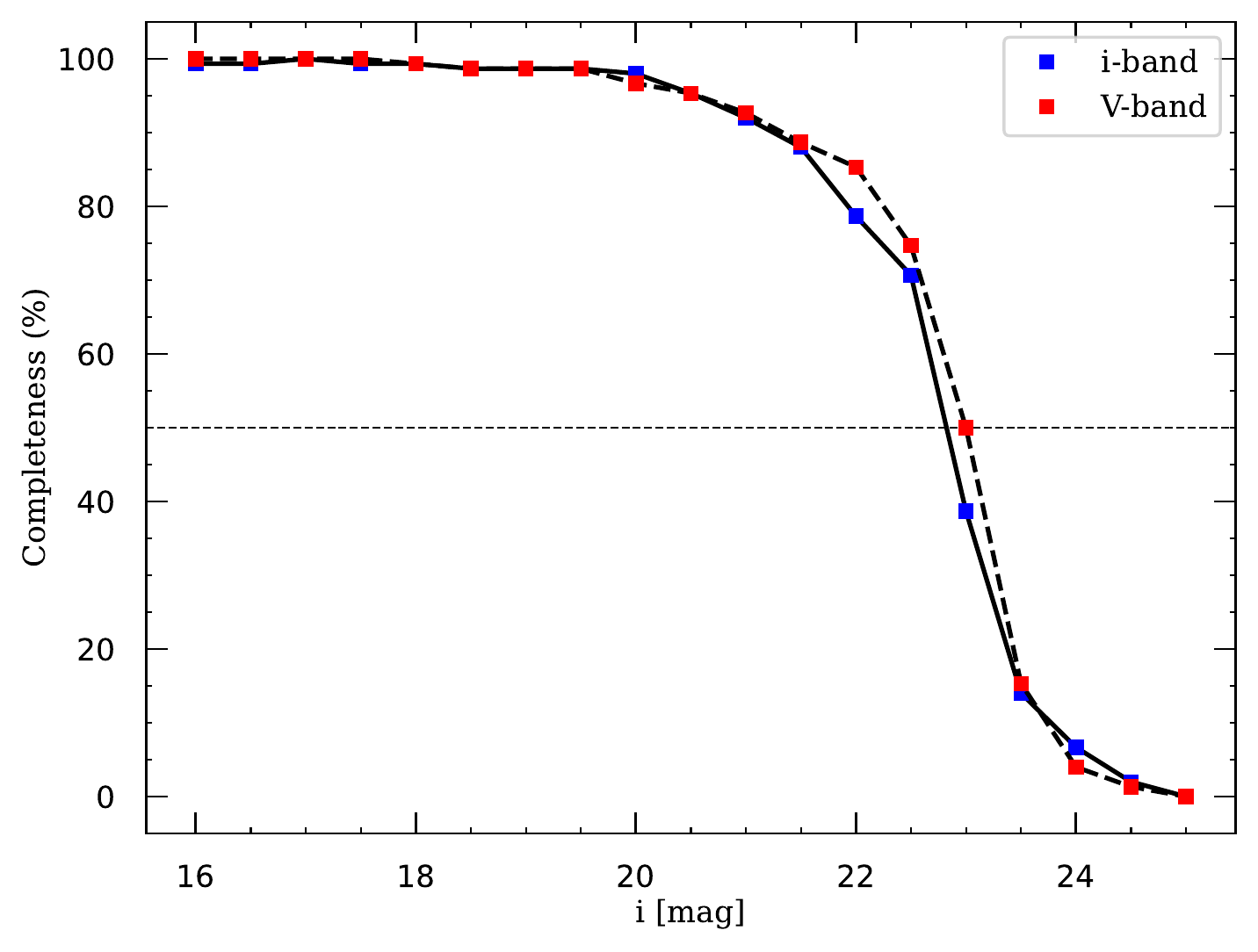}
	\caption{Completeness as a function of $i$ and $V$-band magnitudes. The photometry completeness level is 50$\%$ for stars with $i\approx 23$ mag.}
	\label{fig:Completeness}
\end{figure}

The process of adding 2550 artificial stars was carried out to
evaluate the completeness of the survey, using {\sc ADDSTAR} task \cite[]{1987PASP...99..191S}, in both $i$ and $V$-band single frames, in $17$ discrete $0.5$ magnitude bins started from $16$ to $24.5$ mag. The fraction of recovered artificial stars is estimated by the {\sc ALLFRAME} task. As shown in Fig.~\ref{fig:Completeness}, our survey is sufficiently complete up to $22$ mag in $i$ and $V$-band, near the TRGB (Section \ref{trgb}). Moreover, it is up to $50$\% complete for stars with magnitude $\approx 23$ mag in both filters, which affirms that nearly the entire AGBs and RSGs are detected for our purpose.

\subsection{Calibration}

Different corrections were applied to the stars' magnitude, including aperture correction (by choosing $50$ stars with good photometry to calibrate the PSF-fitting photometry using {\sc DAOGROW} \cite[]{1990PASP..102..932S}, {\sc COLLECT} \cite[]{1993spct.conf..291S}, and {\sc NEWTRIAL} routines), photometric correction (by applying atmospheric extinction correction and zero-points derived from the photometry of standard star images), and relative calibration \cite[]{2020ApJ...894..135S}.
To perform the relative calibration of all images corresponding to one another, we selected $4434$ mutual stars with $i\in [19.0,\,22.0]$ mag. The average magnitude of these chosen stars was taken as the calibration value, and it varies by $-0.01$ to $0.03$ mag.

\begin{figure}[t]
	\centering
	\includegraphics[width=\linewidth, clip]{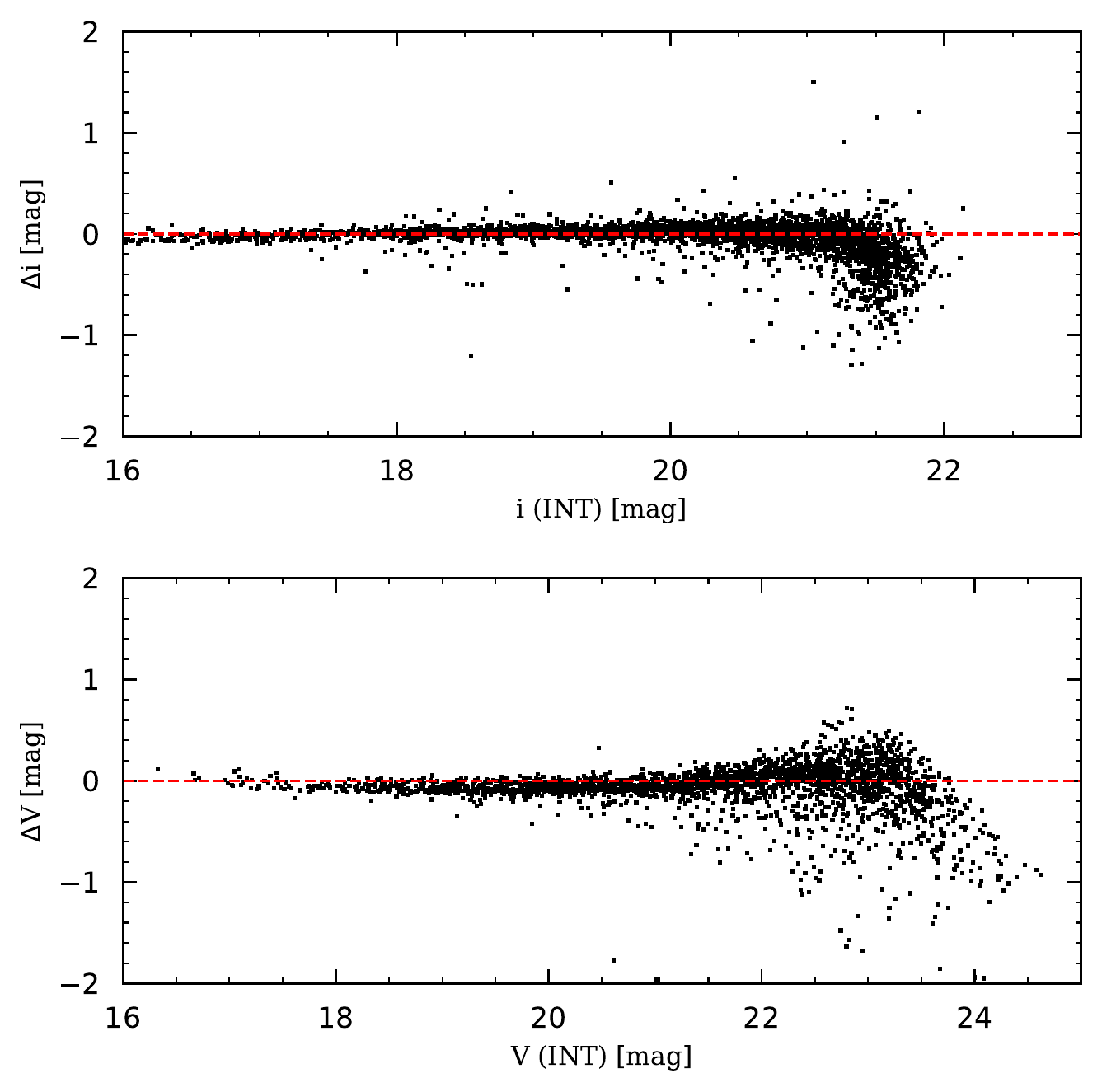}
	\caption{Difference between the magnitude of stars in our catalog and \citet{2016arXiv161205560C} for $i$-band and \citet{2014A&A...570A..78B} for $V$-band.}
	\label{fig:cross}
\end{figure}

In the end, to examine the precision of our estimated magnitudes and the applied calibrations, we cross-matched our catalog with previous surveys. 
Here we present our results for the Pan-STARRS release 1 (PS1) Survey \cite[]{2016arXiv161205560C} in $i$-band and \citet{2014A&A...570A..78B} for $V$-band. Fig.~\ref{fig:cross} shows good accordance up to $i=22$ mag between our photometry and the two other surveys. For fainter stars, differences increase but as far as we are concerned about AGB and RSG stars, our survey is good enough for our purpose.

\section{Variability Analysis}
\label{sec:var}
\begin{figure}[t]
	\centering
	\includegraphics[width=\linewidth, clip]{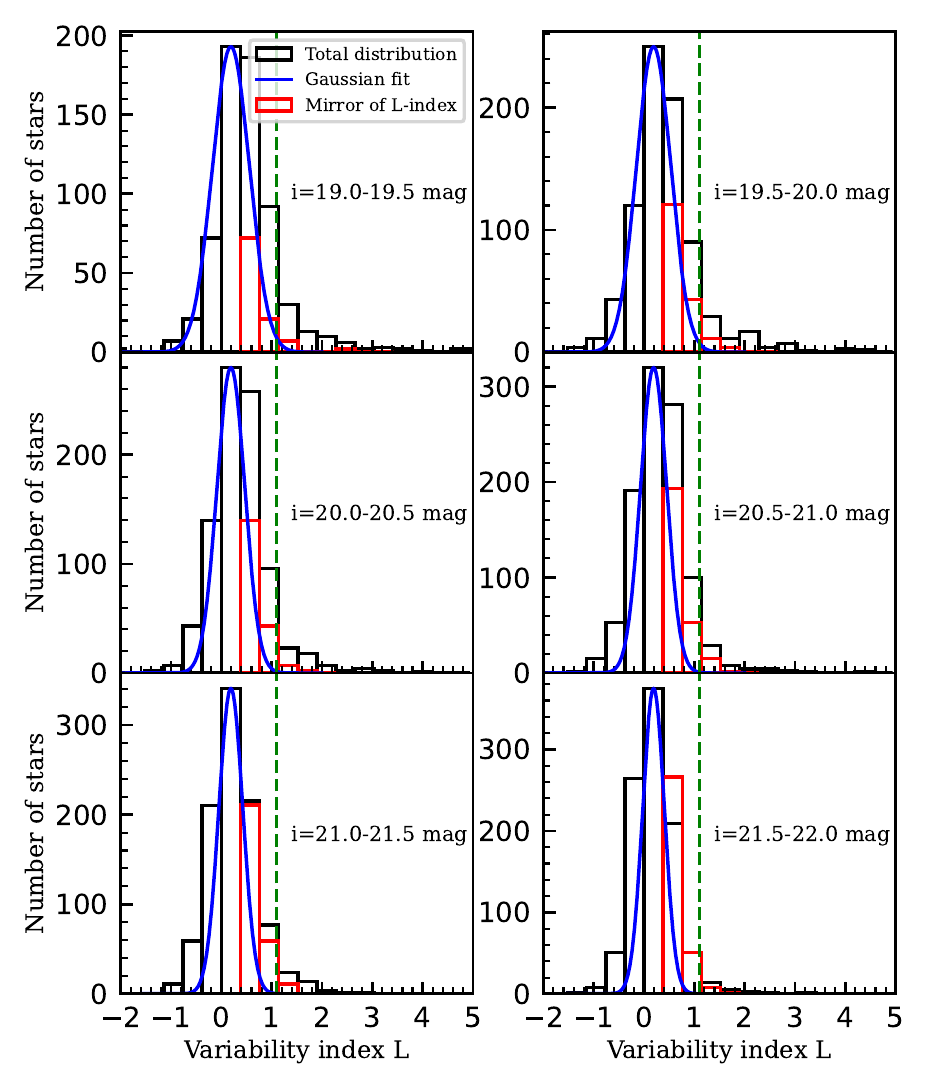}
	\caption{Histogram of the variability index L vs. magnitude for $19$--$22$ mag interval, fitted by the Gaussian function (in blue). The left side of the maximum are non-variable stars, and red bins show the mirror of their distribution. Green dashed lines depict the $L = 1.1$ variability index threshold.}
	\label{fig:hist}
\end{figure}
In order to find LPV stars, we employed the same method presented in \cite{1996PASP..108..851S} to calculate variability index $L$ given in equation \ref{eqn:1}.
\begin{equation}
L = \Bigg(\frac{J\times K }{0.798}\Bigg) \Bigg(\frac{ \sum_{i=1}^{N} w_i }{w_{all}}\Bigg)
\label{eqn:1},
\end{equation}
where the $K$ index is the kurtosis of the magnitude distribution and the $J$ index is defined for $n$ pairs of observations in different or same filters that are on the same night or in a time interval much smaller than the periodicity of LPVs. For variable stars, $J$ has a positive and significant value compared to the non-variables and is calculated as:
\begin{equation}
	J = \frac{\Sigma_{K=1}^{N} w_k \,\textnormal {sign}(P_K)\,\sqrt{|P_K|} }{\Sigma_{K=1}^{N} w_k} 
	\end{equation} where $P_{K}$ is the normalized magnitude residuals of two paired observations \cite[]{1996PASP..108..851S}.
$w$ is the weight parameter set to $1$ for a paired and to $0.5$ for a single observation. $N$ is the number of observations, and $w_{all}$ is the total weight of a star.

Fig.~\ref{fig:hist} shows the histograms of variability index $L$ vs. $i$-band magnitudes in six magnitude bins in a range that covers the AGB-tip and RGB-tip interval. Stars with negative $L$ are non-variables and located at the left side of the maximum ($L \simeq 0$). We expect that non-variable stars form a normal distribution, as they are dominant in number. Hence, we fitted a Gaussian function to the left side of the distribution; the positive part of this fit is assumed to trace the distribution of non-variable stars. We defined a threshold for the variability index $L$ where the ratio of Gaussian fit to actual distribution has diminished to 0.1. It means that $90\%$ of the variable stars in the deviation part of the distribution are selected accurately, and only $10\%$ might be non-variables. We located the variability index threshold at $L = 1.1$, which is shown by vertical green dashed lines in Fig.~\ref{fig:hist}.

Applying $L\geq 1.1$, we found $2539$ LPV candidates in the CCD 4 and $70$ within two half-light radii of SagDIG. We visually examined all the LPV candidates to trace a possible negative correlation with the magnitude changes of a bright neighbor. Fig.~\ref{fig:l} shows the variability index of all sources within two half-light radii of the galaxy vs. stellar magnitude. Genuine LPV candidates are shown as green dots. It is noticeable that they occupy the magnitude interval between AGB-tip and RGB-tip. There are also sources with $L > 1.1$ but marked with black dots as non-variable stars. Nearly all of these sources have magnitudes greater than $20$ mag. They are either foreground contamination (as discussed in Section \ref{sec:cont}) or blended by a bright neighbor, making their estimated index $L$ unreliable.

\subsection{Foreground contamination}
\label{sec:cont}

\begin{figure}[t]
	\centering
	\includegraphics[width=\linewidth, clip]{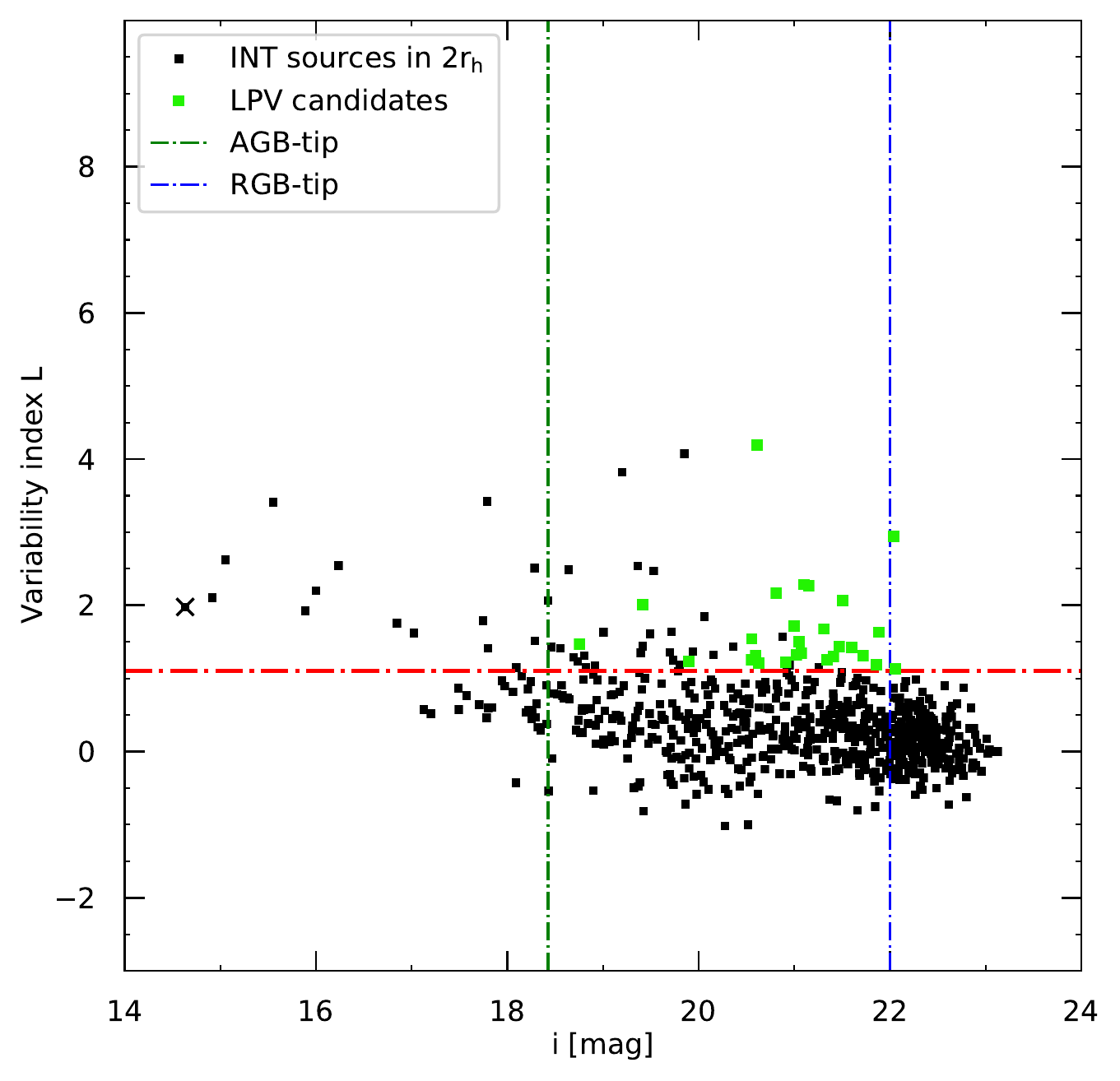}
	\caption{Variability index $L$ vs. $i$-band magnitude. INT sources and LPV candidates within two half-light radii from the center of SagDIG are shown in black and green, respectively. The red horizontal line at $L = 1.1$ indicates the chosen threshold for variable identification. AGB-tip and RGB-tip are shown in green and blue dashed lines, respectively. The black cross is a non-variable source with a large RUWE parameter (see Section \ref{sec:cont}).}
	\label{fig:l}
\end{figure}

SagDIG with {$l$} = $21\rlap{.}^\circ054$ and {$b$} = $-16\rlap{.}^\circ288$ is projected near the center of the MilkyWay, and the observation of the field is negatively influenced by the heavy foreground contamination.
To have a reliable list of LPV candidates, we cross-matched our catalog with the Gaia DR3 \cite[] {2021A&A...649A...1G} and imposed the constraints described in \citet{2020ApJ...894..135S} on the proper motion and parallax of Gaia stars. Gaia DR3 also provides a new parameter RUWE (renormalized unit weight error) that identifies spurious sources.

In total, we found $3418$ foreground stars in CCD 4 and $132$ ones inside $2r_{\rm h}$ from the center of SagDIG, including $15$ stars from our LPV candidates. Among these $15$ stars, only one has a RUWE parameter larger than $1.4$, suggesting it may be a binary star or a star with a disturbing neighbor \cite[] {2021A&A...649A...1G}. We marked it by a cross in Fig.~\ref{fig:l}.

\begin{figure}[t]
	\centering
	\includegraphics[width=\linewidth, clip]{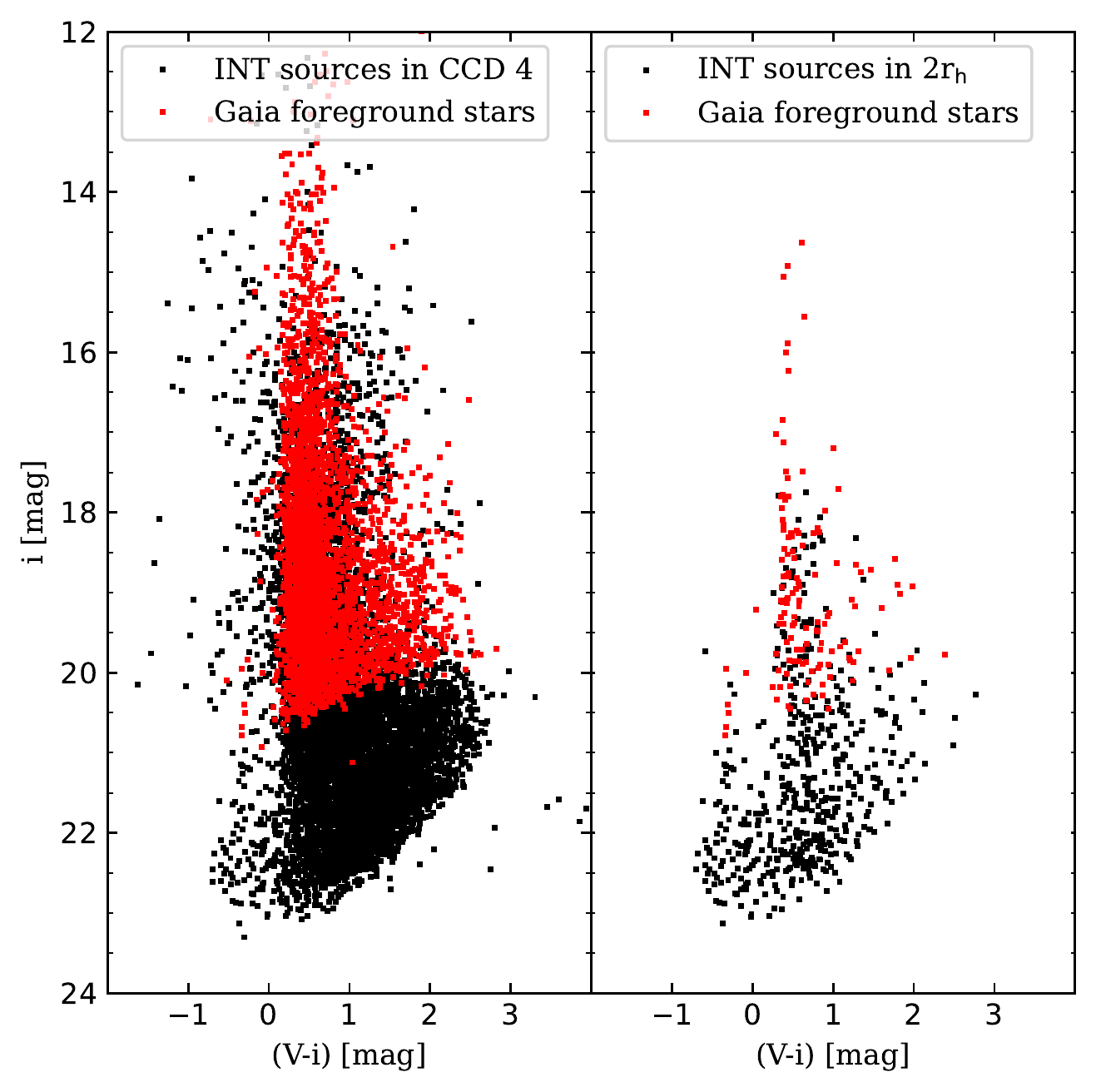}
	\caption{The foreground contamination in line of sight to SagDIG in red dots, obtained by imposing conditions on the Gaia DR3 catalog \cite[] {2021A&A...649A...1G}. Left: stars across CCD 4. Right: stars within $2r_{\rm h}$ from the center of SagDIG.}
	
	\label{fig:gaia}
\end{figure}

\begin{figure}[t]
	\centering
	\includegraphics[width=\linewidth, clip]{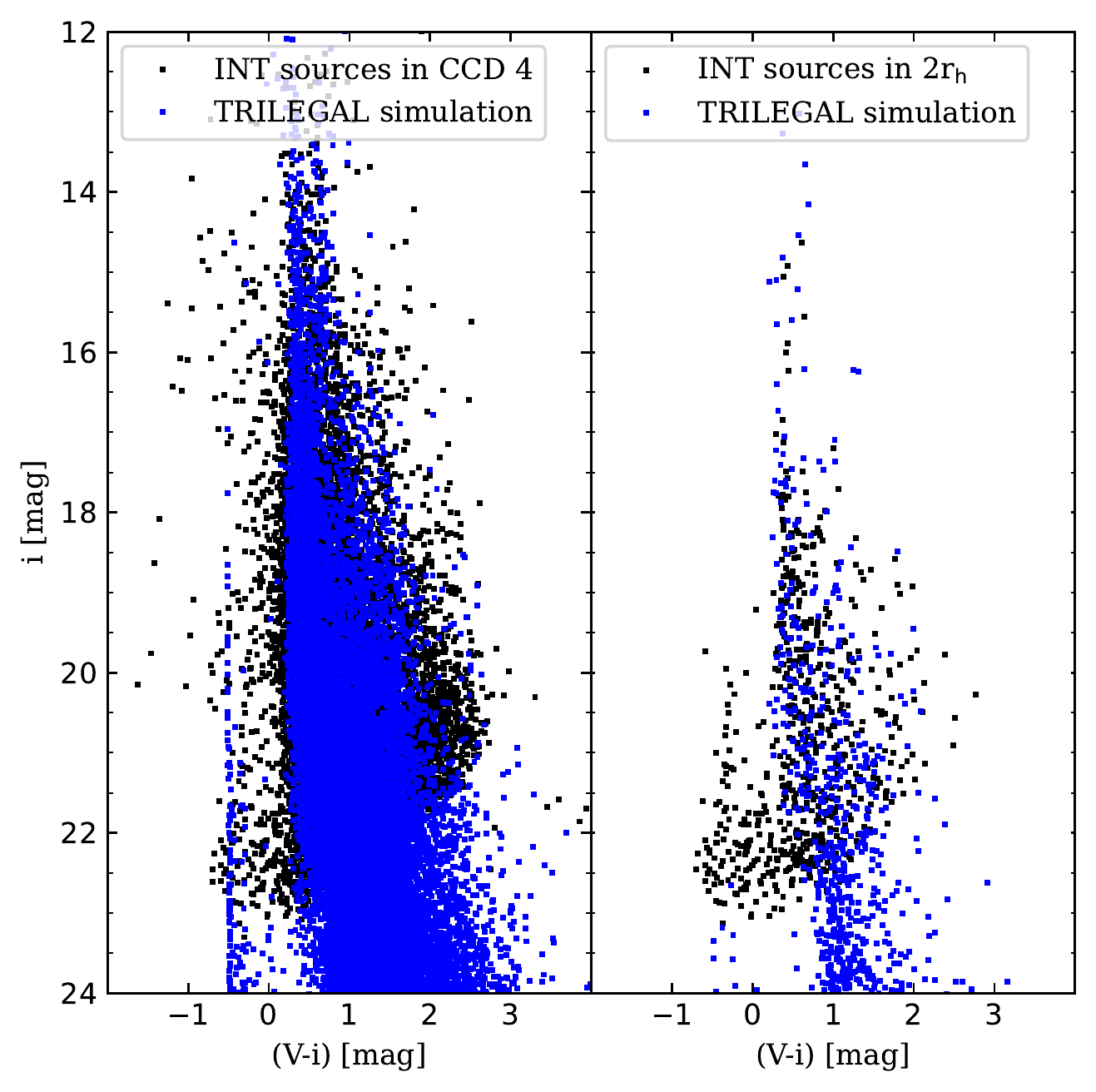}
	\caption{The foreground contamination in line of sight to SagDIG, obtained with TRILEGAL simulation \cite[]{2005A&A...436..895G} represented as blue dots. Left: stars across CCD 4 ($\simeq 0.07$ deg$^2$). Right: stars within $2r_{\rm h}$ from the center of SagDIG ($\simeq 0.002$ deg$^2$).}
	\label{fig:tri}
\end{figure}

As Fig.~\ref{fig:gaia} shows, Gaia DR3 covers only the upper part of the CMD. Hence, it provides no counterparts for stars with a magnitude fainter than $i \simeq 20.5$ mag.
To examine the density of foreground contamination and comparing it to Gaia DR3 coverage, we used the {\sc TRILEGAL} population synthesis code \cite[]{2005A&A...436..895G} with Galactic halo, bulge, thin and thick disk components and applying $V$-band extinction $A_V(\infty) = 0.338$ mag. Fig.~\ref{fig:tri} shows the simulated contamination in the line of sight to SagDIG in blue dots, for the entire CCD 4 ( $\simeq 0.07$ deg$^2$, left) and for $2r_{\rm h}$ from the center of SagDIG ($\simeq 0.002$ deg$^2$, right).
Across the CCD 4 and for $i<20.5$ mag, TRILEGAL estimated $4277$ contamination sources, while Gaia DR3 suggested $3395$. Likewise, Inside $2r_{\rm h}$, TRILEGAL and Gaia DR3 estimated $137$ and $129$ sources, respectively. The estimation ratio shows that Gaia coverage for $i<20.5$ mag is acceptable, especially within $2r_{\rm h}$, which is crucial in our study.

To assess the contamination for $i>20.5$ mag, we compared our catalog and TRILEGAL estimation of the stellar surface density. For this purpose, we only considered stars in an area in the CMD where we expect the LPVs most ($18.4 < i < 22$, $0<V - i<3$ mag), though with a lower magnitude limit of $20.5$ mag. We measured the stellar surface density inside $2r_{\rm h}$ and the entire CCD4 of $0.008$ and $0.003$ arcsec$^{-2}$. In comparison, TRILEGAL estimated the foreground density in these regions at $0.0001$ and $0.004$ arcsec$^{-2}$, respectively. Although the contamination level across the CCD 4 is significant, within $2r_{\rm h}$ and in our considered color and magnitude intervals, it is negligible.

In the end, $27$ stars remained as our final LPV candidates. We omitted $28$ sources with corrupted profiles from the initial list due to blending caused by bright neighbors (see Section \ref{sec:var}), and $15$ stars were among foreground contamination. The photometric properties of the LPV candidates can be found in the appendix.

\subsection{LPV candidates and the Color--Magnitude Diagram}

\begin{figure}[t]
	\centering
	\includegraphics[width=\linewidth, clip]{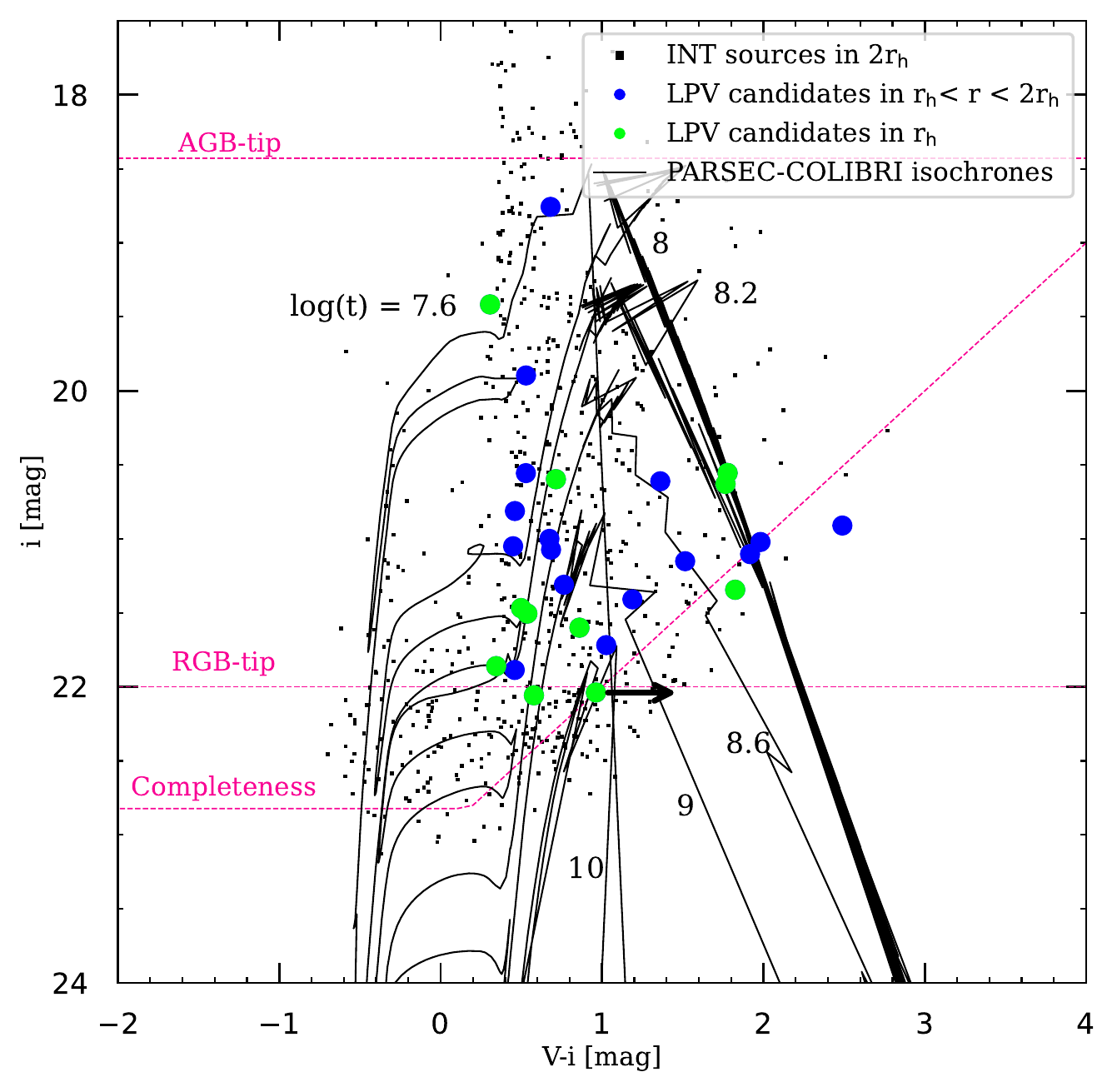}
	\caption{The CMD of stars within $2r_{\rm h}$ of SagDIG along with the PARSEC$-$COLIBRI isochrones. Green dots are LPV candidates inside of $r_{\rm h}=1\rlap{.}^\prime1$, and blue dots are LPV candidates inside $r_{\rm h}<r<2r_{\rm h}$. The AGB-tip, RGB-tip, and completeness limit are shown in magenta dashed lines. The black arrow shows the lower limit for the color of the LPV candidate id $= 5749$.}
	\label{fig:cmdi}
\end{figure}

Fig.~\ref{fig:cmdi} shows the LPV candidates within one and two half-light radii of the galaxy. The PARSEC$-$COLIBRI \cite[]{2017ApJ...835...77M} isochrones with $Z=0.00025$ and our estimated distance modulus in Section \ref{trgb} are overplotted.
The RGB-tip (Section \ref{trgb}) and AGB-tip are presented in magenta dashed lines. Nearly all LPV candidates are brighter than RGB-tip, as we expected.
We estimated the theoretical AGB-tip using the classical core$-$luminosity relation. The absolute bolometric magnitude for a Chandrasekhar core mass is $\approx-7.1$. For SagDIG metallicity and distance modulus, the $\log (t$[Gyr])$=8.0$ isochrone reaches this value at $i=18.43$ mag.

\begin{figure}[t]
	\centering
	\includegraphics[width=\linewidth, clip]{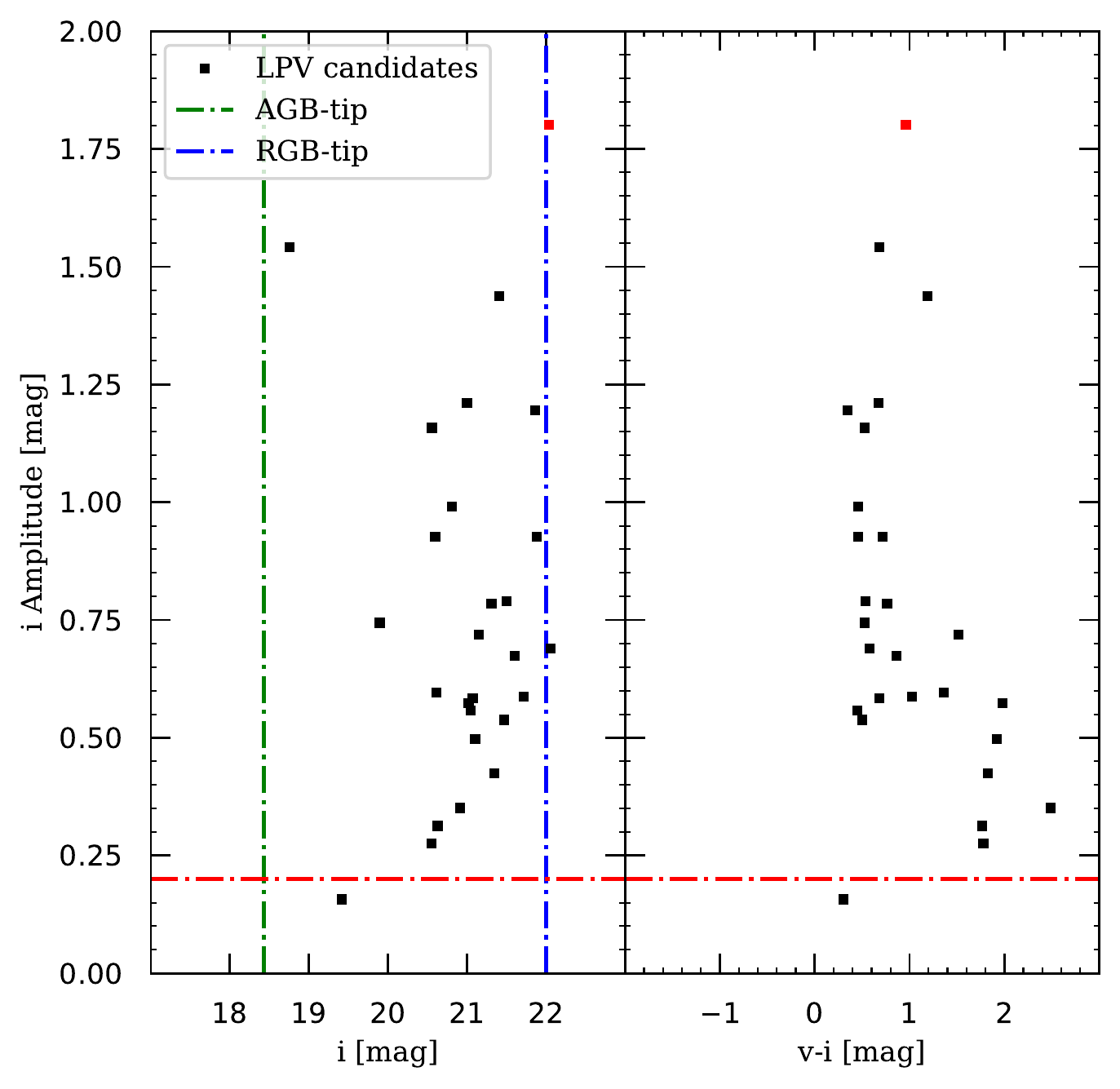}
	\caption{The estimated amplitude of LPV candidates vs. magnitude (left panel) and color (right panel). The red line $A_i=0.2$ mag separates RSGs from AGBs. The blue and green dashed lines in the left panel represent the RGB-tip and AGB-tip, respectively. The red dot has the largest amplitude $A_i = 1.80$ mag, and its light curve is shown in the bottom panel of Fig.~\ref{fig:lcv}.}
	\label{fig:amp}
\end{figure}

The amplitude for a sinusoidal light curve, similar to a variable star's light curve, is obtained by $A =$ $2$$\sigma$$/$$0.707$, where $\sigma$ is the standard deviation of stellar magnitudes. 
As shown in Fig.~\ref{fig:amp}, the amplitude of stars is between $0.16$--$1.8$ mag, and stars with higher amplitude appear dimmer (left panel). The horizontal red line shows a threshold of $0.2$ mag for the amplitude of pulsation. Stars with lower amplitude tend to be RSG stars \cite[]{2011MNRAS.411..263J}.

\begin{figure}[t]
	\centering
	\includegraphics[width=\linewidth, clip]{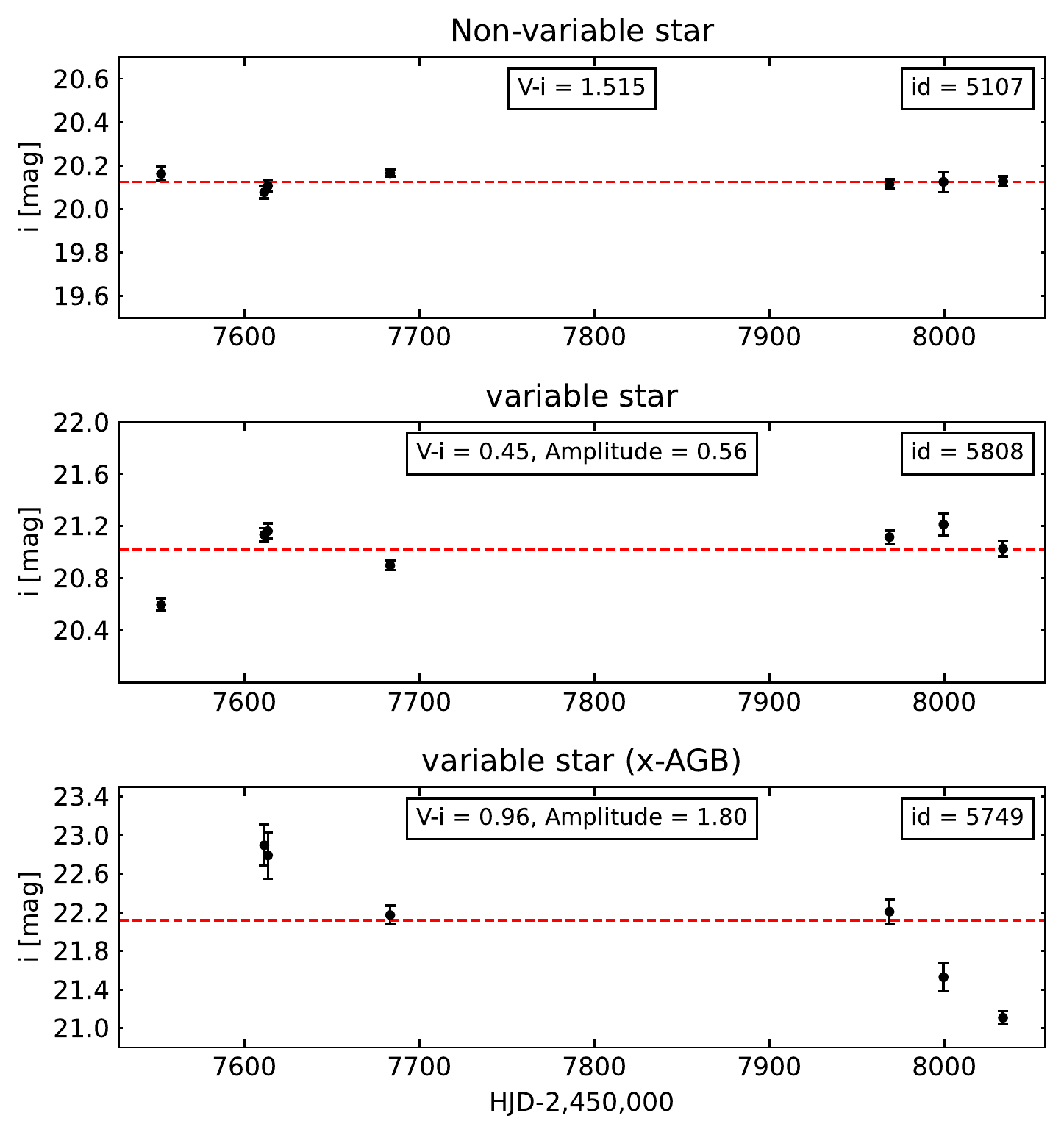}
	\caption{The light curve of a typical non-variable star with the color term (upper panel). The light curve of a variable star with moderate amplitude (middle panel). The Light curve for a variable star (x-AGB; see Section \ref{dusting}) with strong pulsation and high amplitude (bottom panel).}
	\label{fig:lcv}
\end{figure}

In Fig.~\ref{fig:lcv}, the light curves of a non-variable star and two LPV candidates are shown. The id $= 5749$ is an extreme AGB (x-AGB) star (Section \ref{dusting}) and has the largest amplitude ($1.8$ mag). 
Due to limited observation nights, we cannot calculate the period of LPVs precisely, but the color of LPVs correlates with their amplitude and period.
Among our LPV candidates, the id $ = 5749$ is not visible in the $V$-band; hence we considered the completeness limit $50\%$ in the $V$-band as its magnitude. The black arrow in Fig.~\ref{fig:cmdi} indicates this star's lower color term limit. Using Spitzer data in Section \ref{dusting}, we found the color term $i-[3.6]=6.03$ mag for this star, which shows x-AGBs produce a large amount of dust and supports the strong relation of color with amplitude and pulsation.

\section{Distance estimate through locating the tip of RGB}
\label{trgb}

The TRGB is the point where low-mass stars reach their highest luminosity due to the helium flash in their degenerate helium core. At this point, the infrared magnitude (usually I-band) of stars has a slight dependence on metallicity and age \cite[]{1993ApJ...417..553L} and can be used as an extragalactic distance indicator.
TRGB as a standard candle has some advantages over other distance estimators such as RR Lyrae and Cepheid stars. The magnitude of TRGB is brighter than that of RR Lyrae stars, and compared to Cepheids, TRGB is less affected by extinction. Nevertheless, the presence of AGB stars near the tip of RGB can make the TRGB detection problematic in some cases \cite[]{1993ApJ...417..553L}.

\begin{figure}[t]
	\centering
	\includegraphics[width=\linewidth, clip]{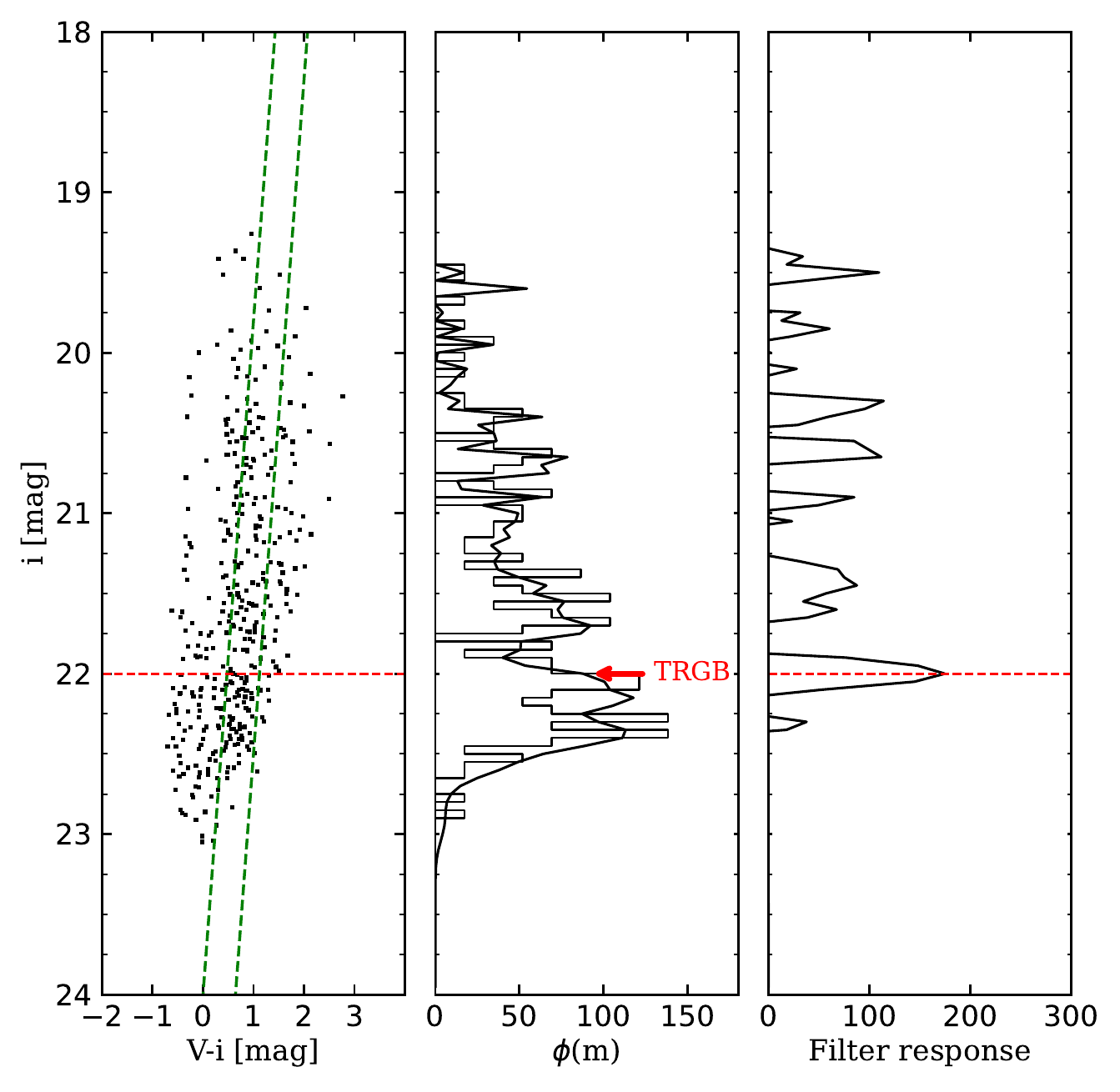}
	\caption{Left panel: The CMD within $2r_{\rm h}$ radii of SagDIG. Stars between green tramlines were employed to detect the TRGB. The dashed red line shows the i-magnitude of TRGB for SagDIG. Middle panel: The smoothed luminosity function $\phi(m)$ overplotted on the luminosity function histogram with binning width $0.05$. Right panel: The output of the edge detection using Sobel kernel.}
	\label{fig:trgb}
\end{figure}

We used the \cite{1996ApJ...461..713S} Gaussian smoothed luminosity function $\phi(m)$ and a Sobel kernel $[-2,-1,0,1,2]$ to perform edge-detection. 
In order to include only RGB stars, we restricted stars within $2r_{\rm h}$ of SagDIG with colors between the green tramlines, as shown in Fig.~\ref{fig:trgb}. We found the $i$-band magnitude of TRGB equal to $22.00 \pm 0.05$ mag for SagDIG (red dashed line in Fig.~\ref{fig:trgb}).
To estimate the distance modulus from $i$-band, we used the luminosity of TRGB in $i$-band $i_0 = 3.53$ mag \cite[]{2016MNRAS.458.1678H} and adopted the i-Sloan correction term $ 2.086 E(B - V)$ with $E(B-V) = 0.124$ mag  \cite[]{1998ApJ...500..525S} to correct the reddening due to the Galactic dust. 
Hence, we found the distance modulus $\mu =$ $25.27 \pm 0.05$ mag.
This value is in good agreement with previous estimates \citep[]{2002A&A...384..393M,2016MNRAS.458.1678H, 2014ApJ...789..147W}.

\section{Star Formation History from LPV candidates}
\label{sec:Star Formation History}

The method we employ to determine the SFH, based on the detection of LPV stars, was first introduced and applied successfully to the M33 galaxy by \citet{8214023}. There are also a number of studies that have applied this method (e.g., IC 1613 -- \citealt{10.1093/mnras/sty3450}; LMC and SMC -- \citealt{2014MNRAS.445.2214R}; NGC 147 and NGC 185 -- \citealt{2017MNRAS.466.1764H}; M33 galactic disc -- \citealt{2017MNRAS.464.2103J}; And I -- \citealt{2020ApJ...894..135S}; And VII -- \citealt{Navabi_2021}). 

In this method, the SFH of a galaxy is defined by the SFR, $\xi(t)$, which is a function of time and describes the amount of star mass formed from gas per year (M$_\odot$ yr$^{-1}$).


\begin{equation}
\xi(t) = 
\frac{\mathrm{d  n^{\prime}(t) } }{\delta t}  
\frac{ \intop\nolimits_{min}^{max} f_{\rm IMF}(m)m\:dm }{ \intop\nolimits_{m(t)}^{m(t+dt)} f_{\rm IMF}(m)\:dm},
\label{eqn:2}
\end{equation}
where $f_{\rm IMF}$ is the initial mass function, defined by \cite{2001MNRAS.322..231K}, with the minimum and maximum stellar masses of 0.02 and 200 M$_\odot$.
$dn^{\prime}(t)$ is the number of LPV stars observed in the age bin $dt$, and $\delta t$ is the pulsation duration (duration of the evolutionary phase of variability).
The numerator shows the total mass of formed stars, and the denominator is the mass of stars formed in the age bin $dt$.




To calculate $\xi(t)$, we need to determine the mass, age, and pulsation duration of LPVs. These stars are at the final stage of evolution and have reached maximum brightness. Hence their luminosity is more directly connected to their birth mass than it is for less evolved stars \cite[]{8214023}.
To convert the i-magnitude to birth mass and estimate $\delta t$, we established relations using PARSEC$-$COLIBRI theoretical stellar isochrones \cite[]{2017ApJ...835...77M}, as explained in \cite{Saremi_2021}. LPVs are assumed to be located at the brightest point of isochrones. For isochrones with the logarithmic age $6.6$ to $10.16$ Gyr, the mass--luminosity relation was determined by fitting the best function to the theoretical i-magnitudes over the associated masses. Likewise, the mass--age relation was obtained by fitting the best function to the distribution of masses and ages. Finally, we estimated the pulsation-duration--mass relation by fitting multiple Gaussian functions to the relative pulsation duration over the mass range.
\cite{Saremi_2021} provides the fits, coefficients, and intercept values of all three relations for different metallicities. We used coefficients and intercepts for SagDIG metallicity after applying the distance modulus and Galactic extinction in line of sight to SagDIG.
As shown in Fig.~\ref{fig:cmdi}, due to the dimming caused by circumstellar dust, LPV candidates with a high reddening are fainter than expected. Hence we applied a correction for all the stars with color $(V-i) > 1.35$ mag. We de-reddened them using the slope of their related isochrones before applying the mass--luminosity relation.

\begin{figure}[t]
	\centering
	\includegraphics[width=\linewidth, clip]{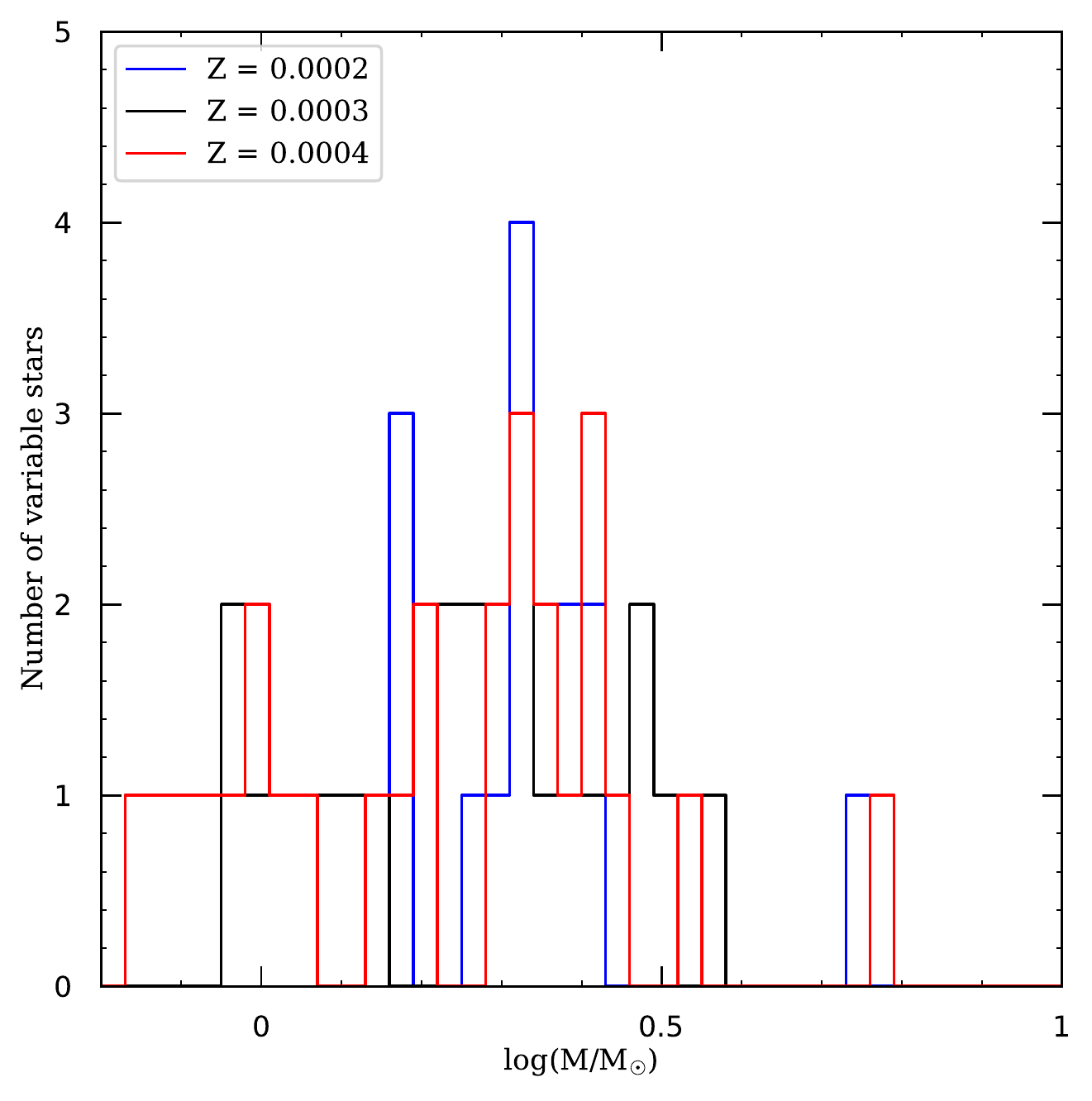}
	\caption{The distribution of stellar masses for three metallicities $Z = 0.0002$ (in blue), $Z = 0.0003$ (in black), and $Z = 0.0004$ (in red). The range of masses shows that all LPV candidates have a low to intermediate mass.}
	\label{fig:range}
\end{figure}

SagDIG's low metallicity has been estimated from RGBs both photometrically ([Fe/H] = $-2.1 \pm 0.2$ dex or $Z = 0.00025$) by  \cite{2002A&A...384..393M} and spectroscopically ([Fe/H] = $ {-1.88_{-0.09}^{+0.13}}$) by \cite{2017ApJ...834....9K}. The metallicity of H\,{\sc ii} regions is evaluated to be [Fe/H]= $−2.07 \pm 0.20$ \cite[]{2002A&A...390...59S}, which shows that the ISM is not considerably metal enriched \cite[]{2002A&A...384..393M}. However,  \cite{2005A&A...439..111M} showed that a slight enrichment of $\sim0.4$ dex, compared to the main population of RGBs is noticeable.
\cite{2017ApJ...834....9K} also showed that the metallicity of stars increases continuously. In this study, we estimated the SFH for three metallicities $Z = 0.0002$, $Z = 0.0003$, and $Z = 0.0004$ ([Fe/H] $\simeq-1.7$), to account for the chemical enrichment. Fig.~\ref{fig:range} shows the mass distribution of the LPV candidates for three choices of metallicities. As we expected for the low mass SagDIG, all LPV candidates have low to intermediate mass.

\section{Results and Discussion }
\label{sec:science}

\subsection{Carbon-rich and Oxygen-rich stars}

\begin{figure*}[ht]
	\makebox[\textwidth][c]{\includegraphics[width=1\textwidth]{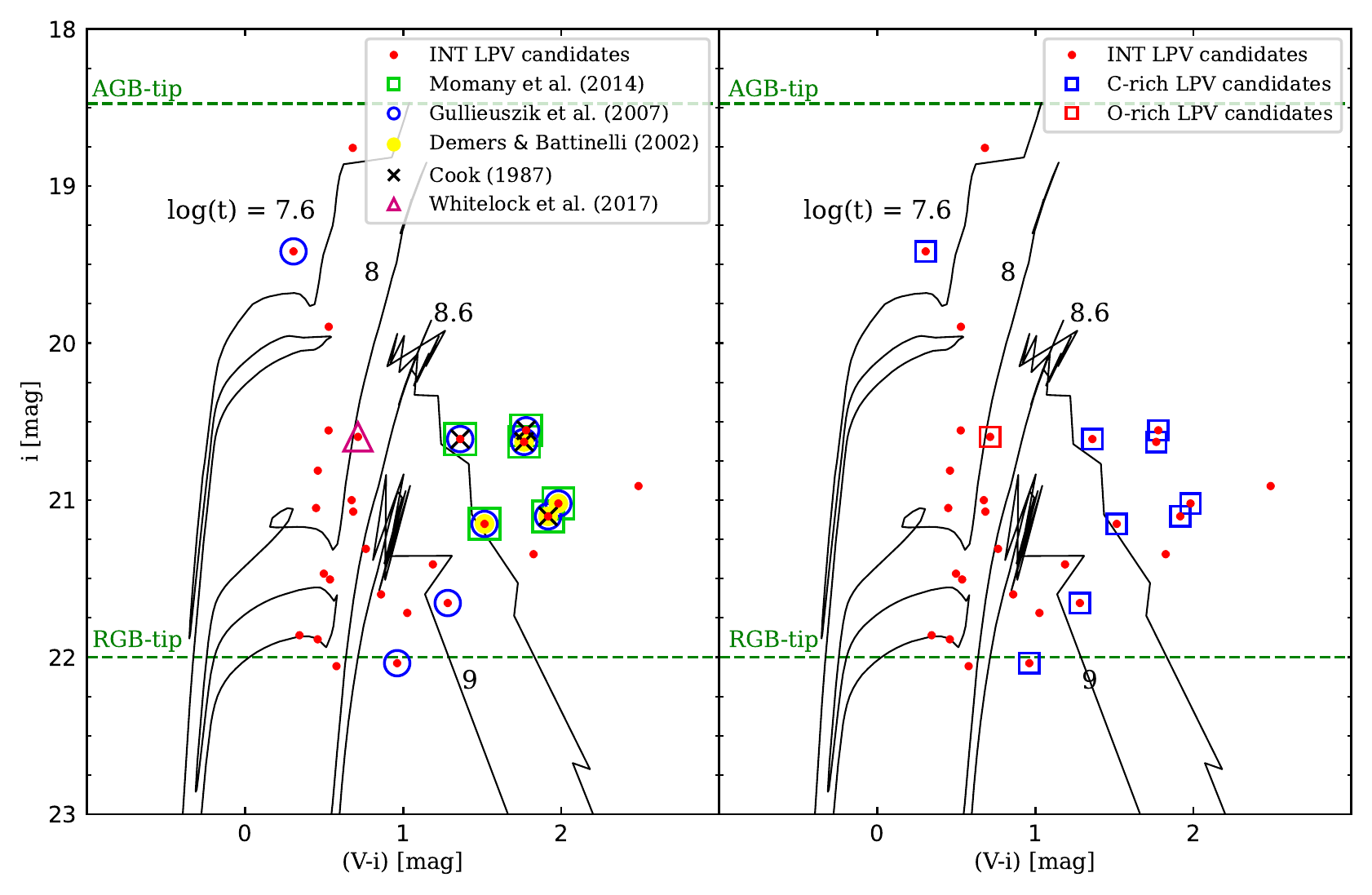}}%
	\caption{CMD of our LPV candidates (red dots) in $i$-band vs $V - i$. Left panel:  Common LPV candidates with other studies are shown. Right panel: C-rich and O-rich stars identified in other studies are marked as blue and red squares.}
	\label{fig:chem_type}
\end{figure*}

With our limited epochs, we are unable to derive the pulsation periods and confirm our LPV candidates. In this section, we compare our list of LPV candidates with other studies that explore the chemical type and variability of the AGB population in SagDIG. This would confirm some of our LPV candidates. Moreover, as part of the objectives of our survey, we are interested in the dust production of LPVs and its relation with their metallicity.

As a low-mass and low luminosity galaxy, SagDIG has few C-rich stars \cite[]{2002AJ....123..238D} in comparison to massive galaxies that harbor hundreds (or more) of them.
\citet{1987PhDT.........6C} was the first who attempted to study SagDIG's C-rich stars, and detected $26$ C-rich stars employing the narrow-band photometry method with filters centered on TiO and CN molecule bands. \citet{2002AJ....123..238D} applied the same approach and found $16$ C-rich stars, eight of which were identified by Cook. For many of Cook's C-rich stars with blue colors, Demers $\&$ Battinelli (D$\&$B) found small CN$-$TiO values and ruled them out.
We found all of the D$\&$B's C-rich stars in our catalog, but we only identified four of them as LPVs. We also found four of Cook's C-rich stars as LPVs; however, two of them are among D$\&$B's C-rich stars. 

\begin{figure*}[ht]
	\makebox[\textwidth][c]{\includegraphics[width=1\textwidth]{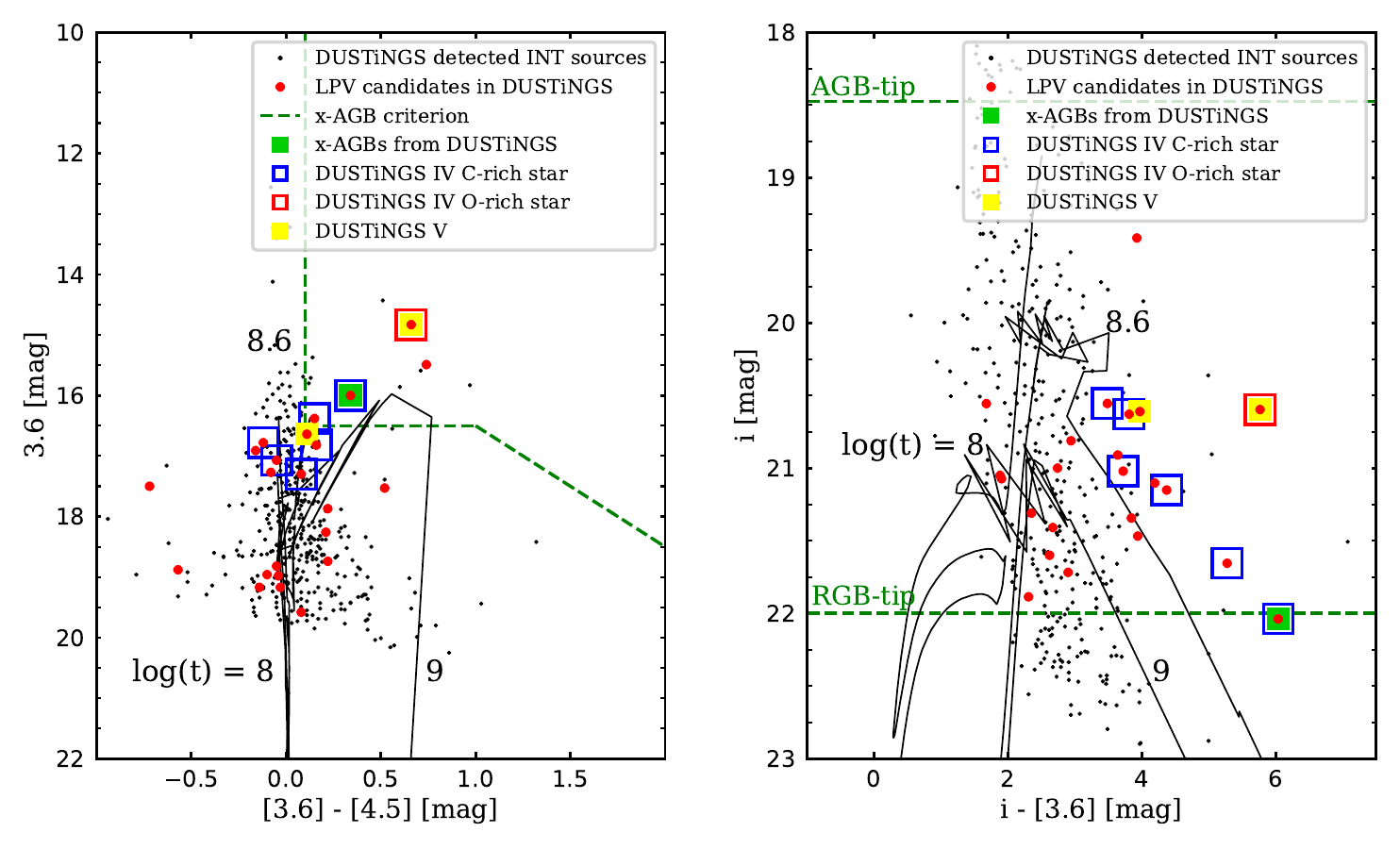}}%
	\caption{CMD of common sources between DUSTiNGS and our catalog in the $[3.6]$-band vs. $[3.6] - [4.5]$ (left panel) and in the $i$-band vs. $i - [3.6]$ (right panel). In both panels, red dots are our LPV candidates. C-rich and O-rich stars from DUSTiNGS are overplotted. The boundary for identifying x-AGBs is shown in the green dashed line in the left panel.}
	\label{fig:spitz}
\end{figure*}

\cite{2007A&A...475..467G} identified $27$ C-rich candidates in SagDIG using broad-band photometry in near-IR. With this approach, more faint and dust enshrouded C-rich stars can be identified compared to narrow-band photometry. They found six new C-rich stars with $J-K_s > 1.9$ mag and $18$ in common with Cook and D$\&$B, which are bluer and mostly have $1.1 < J-K_s < 1.5$ mag.
We found $23$ of their C-rich stars in our catalog and identified nine as LPVs. In addition, one of the C-rich stars in their list is classified as LPV based on our estimated variability index, but it is situated at $2.1 r_{\rm h}$ from the center of SagDIG.
Moreover, Gullieuszik et al. classified three C-rich stars in their catalog as LPVs using ACS/HST \cite[]{2005A&A...439..111M} and D$\&$B's photometry. Two of these LPVs are among our LPV candidates and have a very red color.

\cite{2014A&A...572A..42M} investigated the membership of AGB stars in the Cook and D$\&$B studies by imposing a threshold on the estimated proper motion of these stars. In total, they found $21$ C-rich stars shared with both studies and seven O-rich stars in common with D$\&$B. We identified $24$ stars from Momany et al. list, seven of which are among our LPV candidates. Six of them are C-rich, all with the reddest color.

Fig.~\ref{fig:chem_type} shows the CMD of our LPV candidates in common with other studies (left panel) and their chemical type (right panel). All of the LPV candidates are in the red part of the CMD, and mainly those with the reddest color are C-rich.
In general, exploring the chemical type of our LPVs via other studies leads to one O-rich star detection by \cite{2018MNRAS.473..173W} with a pulsation period of approximately 950 days with a main-sequence mass of $5$ M$_\odot$. The period--luminosity relation of this star reveals that it undergoes Hot Bottom Burning (HBB).

\subsection{X-AGBs and DUSTiNGS survey}

\label{dusting}

Extreme AGBs (x-AGBs) are dusty stars known to be the principal dust producers among thermally pulsating AGBs. Their potentially important role as sources of dust in the first galaxies can be better understood with the help of surveys exploring them in metal-poor galaxies.
DUSTiNGS infrared survey (DUST in Nearby Galaxies with Spitzer -- \citealp{Boyer_2014}) observed $50$ galaxies within $1.5$ Mpc at $3.6$ and $4.5$ $\mu$m. Their purpose was to map the variable AGBs and statistically study this late stage of evolution and the dust production dependence on the parent galaxy's metallicity.

DUSTiNGS used the constraints (green dashed line in left panel Fig.~\ref{fig:spitz}) that had previously been employed for the Magellanic Clouds with more than $90\%$ accuracy \cite[]{2015ApJ...800...51B} to determine x-AGBs.
For SagDIG, which lies in the range of metal-poor galaxies (galaxies with [Fe/H] $<- 2.0$), they found $17$ x-AGBs and six variable x-AGBs. SagDIG has the highest number of very metal-poor x-AGBs among other galaxies in their study. Though it has fewer x-AGBs than massive and more metal-rich galaxies, such stars' existence would suggest no correlation between metallicity and dust production \cite[]{2015ApJ...800...51B}. Four of these x-AGBs are classified as C-rich stars in previous studies, one in D$\&$B and three in  \cite{2007A&A...475..467G}.

In total, we found $397$ common sources with the DUSTiNGS catalog within $2r_{\rm h}$ of SagDIG. 
We identified two x-AGBs in our catalog, but only one of them was classified as a LPV candidate in our study. It is not surprising since x-AGBs are very dusty, and in optical photometry we are solely able to identify those with $[3.6]-[4.5] < 0.5$ mag.
The left panel in Fig.~\ref{fig:spitz} shows the CMD in $3.6$ and $4.5$ $\mu$m with the constraints mentioned earlier. Most of our LPVs lie on the $\log(t)= 8$ and $8.6$ Gyr isochrones and have $[3.6]-[4.5]<0.5$ mag. There are four LPV candidates within x-AGBs criterion. The right panel in Fig.~\ref{fig:spitz} shows the distribution of LPVs in the $i$ and $3.6$ $\mu$m CMD. The confirmed x-AGB star (the green square) is faint and dusty.

For further investigation of the chemical types of dusty AGBs in very metal-poor galaxies including SagDIG, DUSTiNGS IV \cite[]{2017ApJ...851..152B} used additional observations from WFC3/IR with F127M, F139M, and F153M filters (best matched to H$_2$O, CN, and C$_2$ features). These filters are more efficient than optical narrow-band and near-IR broad-band photometry in distinguishing between C-rich and O-rich stars and identifying very faint dust-enshrouded AGBs.
For SagDIG, they found $37$ AGBs, most of them classified as C-rich stars. We identified six C-rich stars (one of them is the x-AGB from DUSTiNGS II and the others were among C-rich stars in previous studies). They identified two very dusty O-rich stars. We identified both of these stars, but only one of them is among our LPV candidates, which is the same O-rich star as discovered by \cite{2018MNRAS.473..173W}. As can be seen in Fig.~\ref{fig:spitz}, the O-rich star (red square) has a large $i - [3.6] = 5.8$ mag, comparable to the C-rich x-AGB star (green square).
This is an indication that, unlike theoretical models prediction for low metallicity environments, O-rich stars produce a large amount of dust consistent with $[3.6]-[4.5]$ color.

In DUSTiNGs V, \citet{2019ApJ...877...49G} used extra epochs of observations with the $Spitzer$ telescope to investigate the period--luminosity (P--L) relation in galaxies with different metallicity. They showed that the P--L relation is unchanged at low metallicity, and the dust production and pulsation are linked as previous studies suggested.
They estimated the period of three LPVs in SagDIG. As shown in Fig.~\ref{fig:spitz}, two of these stars are among our LPV candidates. One of them has a period of $928$ days in $3.6$ $\mu$m and $\sim2000$ days in $4.5$ $\mu$m. The other is the most metal-poor O-rich AGB star ever known (the same as Whitelock et al. found) and is currently experiencing HBB. They estimated the period of the O-rich LPV to be $\sim2000$ days. Since the estimation was made with insufficient data, the $950$ days that Whitelock et al. estimated is more reliable.

\subsection{The SFH}
\label{sfh_sagdig}

\begin{figure}[t]
	\centering
	\includegraphics[width=\linewidth, clip]{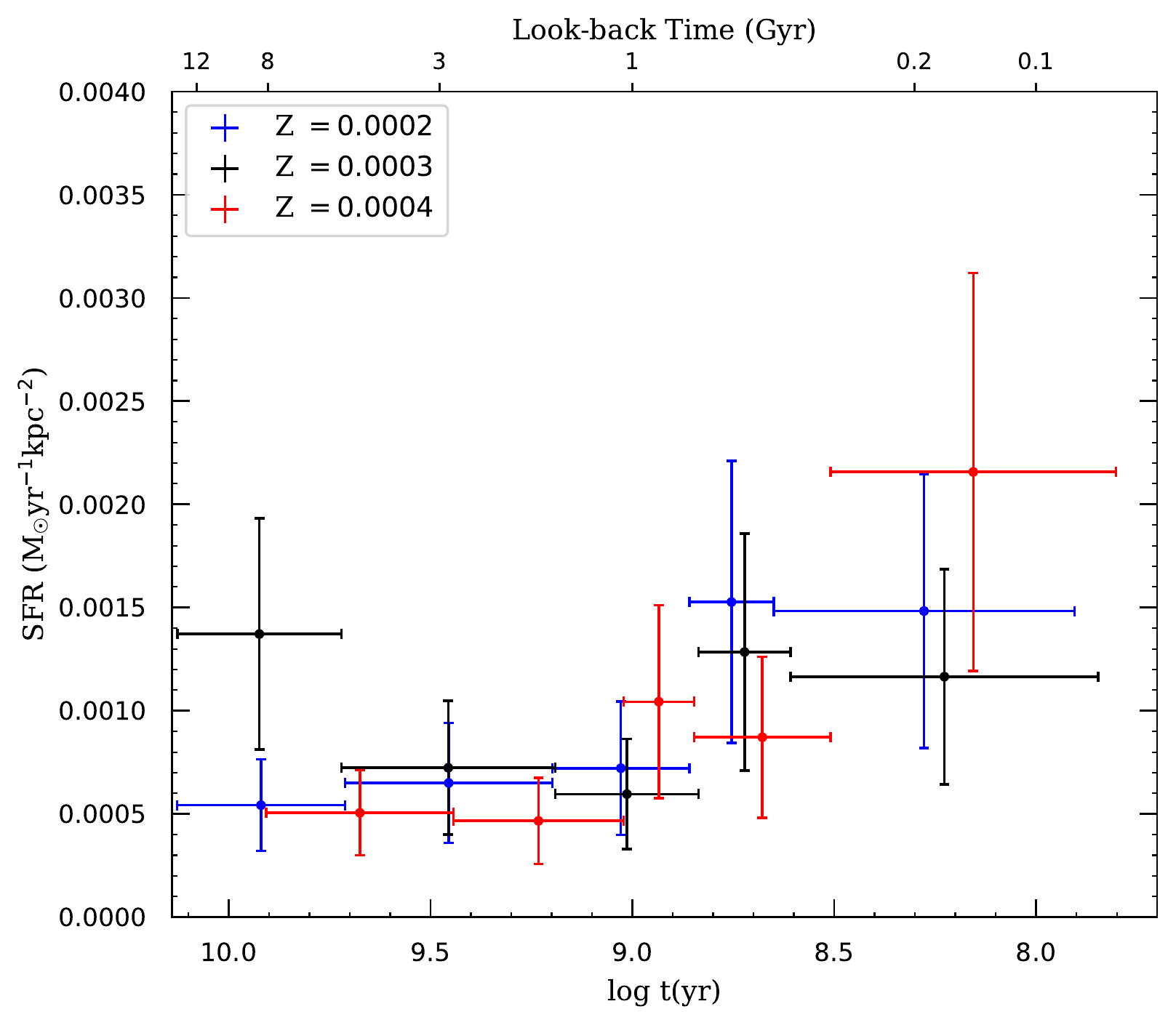}
	\caption{The SFR of SagDIG within $2r_{\rm h}$ for three metallicities $Z = 0.0002$ (in blue), $Z = 0.0003$ (in black), and $Z = 0.0004$ (in red).}
	\label{fig:sfr}
\end{figure}

To construct the SFH of SagDIG, we defined different age bins consisting of nearly an equal number of LPV candidates to have a uniform uncertainty, and we used the method explained in Section \ref{sec:Star Formation History}. Fig.~\ref{fig:sfr} represents the SFR within $2r_{\rm h}$ ($0.68$ kpc$^2$) for three metallicities, starting around $13$ Gyr ago ($\log$($t$[Gyr]) $= 10.13$) up to $63$ Myr ago ($\log$($t$[Myr]) $= 7.8$).
The vertical error bars denote SFH errors in each age bin based on Poisson statistics. As shown in Fig.~\ref{fig:sfr}, SagDIG has had a continuous star formation activity over its lifetime with different rates and has experienced an enhancement of star formation since $\simeq 1$ Gyr ago, which is quite common among dwarf irregular galaxies \cite[]{2014ApJ...789..147W}.

Assuming $Z = 0.0002$, SagDIG experiences the lowest SFR $\simeq 0.0005 \pm 0.0002$ M$_{\odot}$yr$^{-1}$kpc$^{-2}$ around $13$ Gyr ago ($\log$($t$[Gyr]) $= 10.13$--$9.7$). Considering the error bars, the SFR remains nearly constant up to $\simeq$ $0.6$ Gyr ago ($\log$($t$[Gyr] $\simeq 8.8$). Afterward, the SFR increases and its peak reaches $\simeq 0.0015 \pm 0.0007$ M$_{\odot}$yr$^{-1}$kpc$^{-2}$ at $0.6$--$0.08$ Gyr ago ($\log$($t$[Gyr])$\simeq 8.8$--$7.9$), which is around $3$ times higher than SagDIG's lowest SFR. 
In general, the SFH for all three metallicity choices does not differ considerably, except for the younger population.
Assuming $Z = 0.0004$, the peak of SFR increases to $0.0021 \pm 0.0010$ M$_{\odot}$yr$^{-1}$kpc$^{-2}$ and shifts toward younger ages ($\log$($t$[Gyr]) $\simeq 8.5$--$7.8$).

Considering the errors, our result is in good accordance with \cite{1999A&A...352..363K}. They compared a synthetic CMD to $I$ and $V$-band observations from the $2.5$ m Nordic Optical Telescope. They estimated the SFR $0.00028 \pm 0.00001$ and $0.00144 \pm 0.00017$ M$_{\odot}$yr$^{-1}$kpc$^{-2}$ for the age range $15$--$0.2$ Gyr and $0.2$--$0.05$ Gyr, respectively. We note that we estimated a higher value of SFR at $1$--$0.2$ Gyr ago. However, as \cite{1999A&A...352..363K} mentioned, their estimation at this interval is uncertain due to poor sampling caused by foreground stars and low star counts in the CMD. They also adopted a metallicity range of $0.0004 < Z < 0.0005$ for the reason that the stellar evolutionary models were incomplete for lower metallicities at that time. Moreover, the considered area in their study was around $0.46$ kpc$^{-2}$ and smaller than our field. 
\cite{2005A&A...439..111M} estimated the SFR $\simeq 0.00027$ M$_{\odot}$yr$^{-1}$ for the last $0.6$ Gyr, in a smaller field, using HST/ACS images and the luminosity function of main sequence stars. Their result was in good agreement with \cite{1999A&A...352..363K}, though, for the last $0.02$ Gyr, their estimation is lower and more compatible with H$_{\alpha}$ flux from \cite[]{1991ApJ...383..148S}.

Our study only covers SFH up to $63$ Myr ago. 
However, other studies demonstrated the rapidly growing state of SagDIG by tracing blue stars. For example, \cite{2012AJ....144..134H} used UV flux from the GALEX survey, and \cite{1991ApJ...383..148S} and \cite{2012AJ....144..134H} used H$_{\alpha}$ flux of ionized hydrogen from the only H\,{\sc ii} region in SagDIG. UV flux is a good tracer of the SFR over the $100$--$200$ Myr ago, whereas H$_{\alpha}$ flux only represents the star formation in the past $10$ Myr.
The estimated SFR in the last age bin in our study is very close to the calculated SFR from far-UV images of the GALEX survey \cite[]{2012AJ....144..134H}. They found $\log$ SFR$^{FUV} _ D = - 2.11 \pm 0.01$ M$_{\odot}$ yr$^{-1}$ kpc$^{-2}$. The value is scaled
to the area with disc scale length $R_{D}$ = $0.23 \pm 0.03$ kpc. By multiplying the area, the SFR is equal to $0.0013$ M$_{\odot}$ yr$^{-1}$, which is close to our estimation for the latest age interval.

For $Z = 0.0002$, $Z = 0.0003$, and $Z=0.0004$ we estimated the total stellar mass M$_*(<2r_{\rm h})=$ ($5.4 \pm 2.3$) $\times$ $10^ 6$ M$_{\odot}$, ($9.6 \pm 4.0$) $\times$ $10^ 6$ M$_{\odot}$ ,and ($3.0 \pm 1.3$) $\times$ $10^ 6$ M$_{\odot}$, respectively. This is in good agreement with previous estimates. \cite{2012AJ....144....4M} determined M$_{*} = 3.5 \times 10^ 6$ M$_{\odot}$ with the assumption of stellar mass to light ratio of $1$. \cite{1997ApJ...490..710Y} estimated ($2.5$--$7.5$)$\times 10^6$ M$_{\odot}$, assuming stellar mass to light ratio of $1$--$3$.

\begin{figure}[t]
	\centering
	\includegraphics[width=\linewidth, clip]{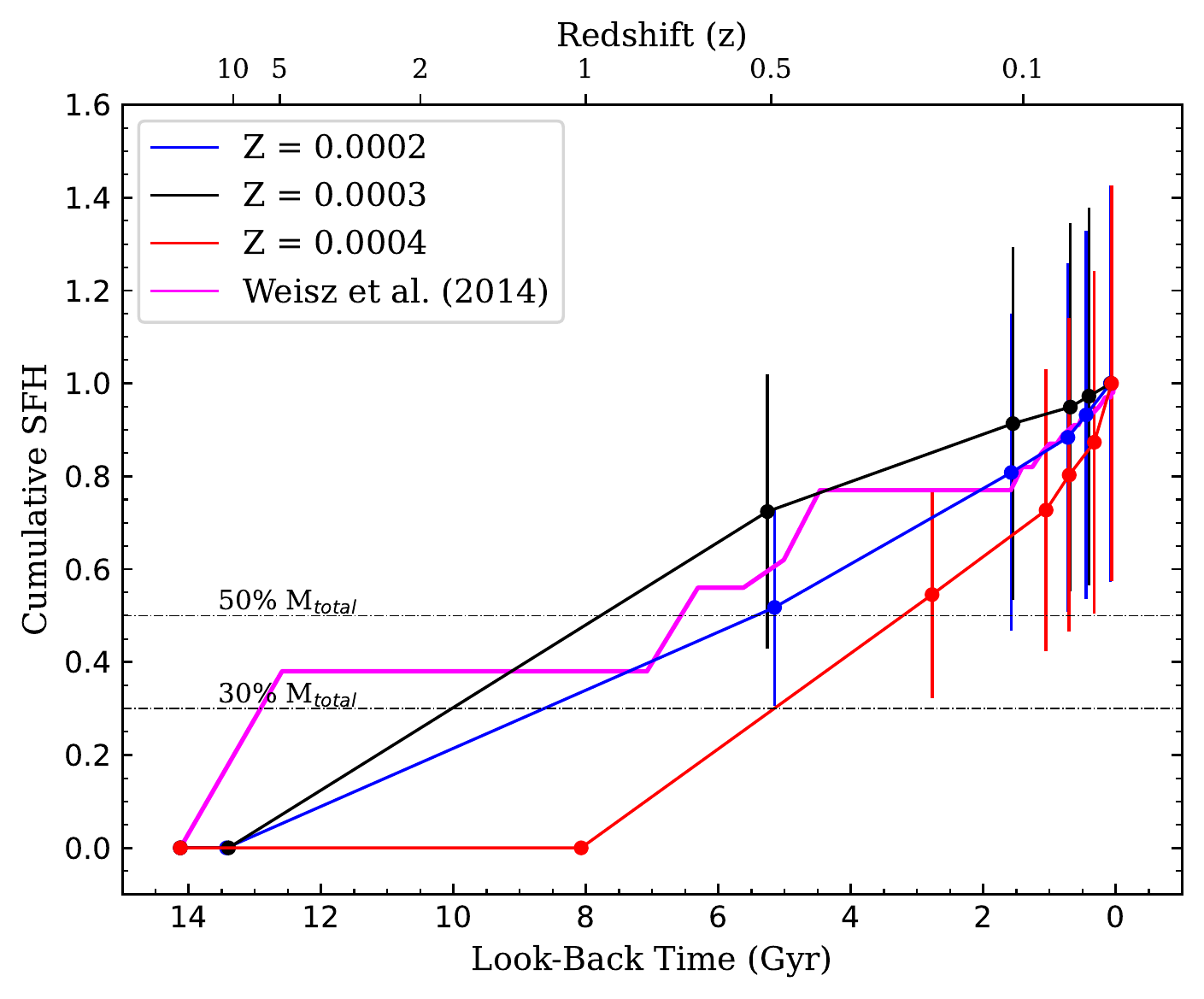}
	\caption{Cumulative SFH of SagDIG within $2r_{\rm h}$ for three metallicities $Z = 0.0002$ (in blue), $Z = 0.0003$ (in black), and $Z = 0.0004$ (in red). The black dashed lines represent $30\%$ and $50\%$ of the total stellar mass formed. The cumulative SFH from \cite{2014ApJ...789..147W} is overlaid to compare.}
	\label{fig:cum}
\end{figure}

To further study the amount of formed mass in different epochs, we defined the cumulative SFH as the fraction of stellar mass formed as a function of look-back time. To better compare our results with \cite{2014ApJ...789..147W}, which used HST deep images to obtain the cumulative SFH of LG dwarf galaxies, we overlay their measured SFH in Fig.~\ref{fig:cum}. They found that, like most dwarf irregulars, SagDIG undergoes an increase in star formation since $z = 1$. We also found an increase in the SFH since $z \simeq 1$ ($\log$($t$[Gyr]) $\simeq 8.8$--$9.0$), though it is better noticeable in Fig.~\ref{fig:sfr}.
As shown in Fig.~\ref{fig:cum}, our results imply that SagDIG formed only $30\%$ of its total stellar mass prior to $\simeq 8.5$ Gyr ago ($z = 1.2$), and the fraction increased to about $50\%$ by 5 Gyr ago. \cite{2014ApJ...789..147W} estimated that $30\%$ of SagDIG's stellar mass formed before $z = 2$ ($10.5$ Gyr ago), and around $50\%$ formed before $6$ Gyr ago and concluded the latter as the mean age of SagDIG.

\subsection{Spatial distribution and age gradient}

\begin{figure}[t]
	\centering
	\includegraphics[width=\linewidth, clip]{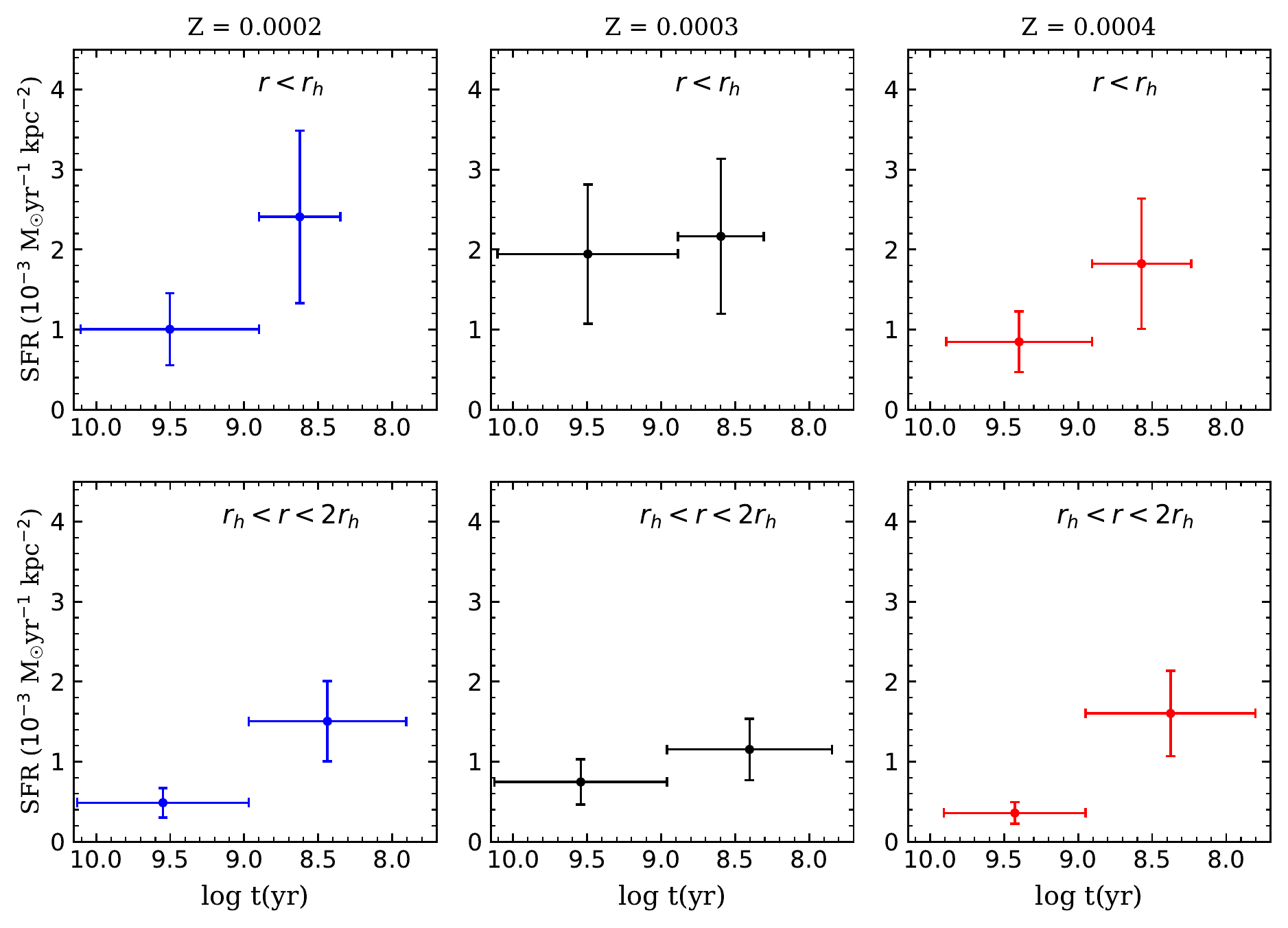}
	\caption{The SFR of SagDIG for three metallicities $Z = 0.0002$, $Z = 0.0003$, and $Z = 0.0004$ inside $r_{\rm h}$ and within $r_{\rm h}$--$2r_{\rm h}$.}
	\label{fig:sfr-rh}
\end{figure}

To investigate the age gradient in SagDIG, we divided the LPV candidates into two groups based on their distance from the galaxy's center. We estimated the ratio of SFR in older times ($\log t$(yr)$>9.2$, $Z = 0.0002$) to that of more recent times ($\log t$(yr)$<9$, $Z = 0.0004$), for both $r<r_{\rm h}$ and $r_{\rm h} < r < 2r_{\rm h}$, to be $0.56$ and $0.31$, respectively. 
Considering the errors (minimum of old SFR and maximum of recent SFR), the ratios become $0.23$ and $0.15$.
The slightly higher ratio of SFR in the inner region and the extension of SFR to more recent times in the outer part, as shown in Fig.~\ref{fig:sfr-rh}, could suggest an inside-out formation scenario by internal or external processes.
However, the difference in the ratio of SFR between the two regions is not significant enough to strongly confirm this scenario.
Moreover, considering the isolation of SagDIG, any external interaction or merger with another galaxy is highly unlikely.

The evolution of dwarf galaxies is susceptible to their environment. Although SagDIG is an isolated galaxy today, could it have experienced some interactions in the past that would have caused the spatial extension of the intermediate--age population?  
Our analysis here is limited to $2.2$ arcmins. Therefore, any deduction from our results about the structure and the possibility of tidal interactions might not be robust.
Nevertheless, the extended structure of SagDIG has been studied out to 5 arcmins \cite[]{2014A&A...570A..78B} and beyond that \cite[]{2016MNRAS.458.1678H}, suggesting the possibility of tidal interaction in the past.
SagDIG's H\,{\sc i} has a distributed morphology and a large gradient in the kinematics, though with no sign of rotation. Aside from possible internal processes, \cite{2014A&A...570A..78B} mentioned that tidal interaction may have played a role in the characteristics of H\,{\sc i}.
\cite{2016MNRAS.458.1678H} found a mild distortion in RGB distribution in the exterior part of SagDIG along the major axis. They speculated it is a sign of tidal interaction in the past with a small object like a globular cluster. However, they also offered another explanation and found it more likely: SagDIG has a secondary component that encompasses the main body, such as a stellar halo that extends beyond 5 arcmins.

By investigating the spatial distribution of the intermediate–age and old populations, \cite{2017ApJ...834...78M} found that in SagDIG, TP-AGBs have a longer scale length in comparison to RGBs.
They also found breaks in the RGB and AGB surface density at $\lesssim 3^\prime$, where the H\,{\sc i} column density also decreases.
They concluded that their findings, in addition to the distributed morphology and the kinematics of H\,{\sc i}, suggest that SagDIG has undergone tidal interactions

Considering the tidal scenario and the uncertain durability of tidal interaction in a low-mass system, the current position of the intermediate-age population does not reflect the formation scenario of these populations; whether they were pulled out after formation or formed in the outer extremities at the first place \cite[]{2017ApJ...834...78M}.

\section{Summary and Conclusions}
\label{sec:conclusions}
This paper presented the star formation history of the SagDIG galaxy based on detecting LPVs in $i$- and $V$-band observations taken during 2016--2017 with the Isaac Newton Telescope. 
$27$ LPV candidates were detected within two half-light radii of SagDIG; ten were in common with other studies, including $9$ C-rich stars and one O-rich star.
Considering a slight metallicity enrichment (Z = 0.0004), we estimated the SFR $0.0021 \pm 0.0010$ M$_{\odot}$yr$^{-1}$kpc$^{-2}$ at $0.3$--$0.06$ Gyr ago ($\log$($t$[Gyr]) $\simeq 8.5$--$7.8$) that is $\simeq 4$ times higher than the lowest SFR in older ages with Z=0.0002.
Moreover, SagDIG assembled $30\%$ and $50\%$ of its stellar mass before $8.5$ and $5$ Gyr ago, respectively.
SagDIG has had continuous star formation, and like many dIrrs, its SFR has been increased since $z \simeq 1$.
The total stellar mass within $2r_{\rm h}$ from the center of SagDIG was estimated for three different choices of metallicity. For $Z=0.0002$ and $Z=0.0004$ we estimated the stellar mass ($5.4 \pm 2.3$) $\times$ $10^ 6$ and ($3.0 \pm 1.3$)$\times$ $10^ 6$ M$_{\odot}$, respectively.
We also determined a distance modulus of $\mu = 25.27 \pm 0.05$ mag using the TRGB method.

\section*{Acknowledgments}
The observing time for this survey was provided by the Iranian National Observatory and the UK-PATT allocation of time to programs I/2016B/09 and I/2017B/04 (PI: J. van Loon).
The authors thank the Iranian National Observatory and the School of Astronomy (IPM)
for the financial support of this project. We are
grateful to Peter Stetson for sharing his photometry routines.
We thank James Bamber, Rosa Clavero, Arash Danesh, Ghassem Gozaliasl, Alireza Molaeinezhad, Mojtaba Raouf, Philip Short, and Lucia Su\'arez-Andr\'es for their help with the observations. Finally, we are grateful to the anonymous referee for carefully reading the manuscript and for helpful comments and suggestions, which helped us to improve the quality of the manuscript.

\bibliography{sample631}{}

\begin{thebibliography}{}
\expandafter\ifx\csname natexlab\endcsname\relax\def\natexlab#1{#1}\fi
\providecommand{\url}[1]{\href{#1}{#1}}
\providecommand{\dodoi}[1]{doi:~\href{http://doi.org/#1}{\nolinkurl{#1}}}
\providecommand{\doeprint}[1]{\href{http://ascl.net/#1}{\nolinkurl{http://ascl.net/#1}}}
\providecommand{\doarXiv}[1]{\href{https://arxiv.org/abs/#1}{\nolinkurl{https://arxiv.org/abs/#1}}}

\bibitem[{{Beccari} {et~al.}(2014){Beccari}, {Bellazzini}, {Fraternali},
  {Battaglia}, {Perina}, {Sollima}, {Oosterloo}, {Testa}, \&
  {Galleti}}]{2014A&A...570A..78B}
{Beccari}, G., {Bellazzini}, M., {Fraternali}, F., {et~al.} 2014, \aap, 570,
  A78, \dodoi{10.1051/0004-6361/201424411}

\bibitem[{Boyer {et~al.}(2014)Boyer, McQuinn, Barmby, Bonanos, Gehrz, Gordon,
  Groenewegen, Lagadec, Lennon, Marengo, Meixner, Skillman, Sloan, Sonneborn,
  van Loon, \& Zijlstra}]{Boyer_2014}
Boyer, M.~L., McQuinn, K. B.~W., Barmby, P., {et~al.} 2014, The Astrophysical
  Journal Supplement Series, 216, 10, \dodoi{10.1088/0067-0049/216/1/10}

\bibitem[{{Boyer} {et~al.}(2015){Boyer}, {McQuinn}, {Barmby}, {Bonanos},
  {Gehrz}, {Gordon}, {Groenewegen}, {Lagadec}, {Lennon}, {Marengo}, {McDonald},
  {Meixner}, {Skillman}, {Sloan}, {Sonneborn}, {van Loon}, \&
  {Zijlstra}}]{2015ApJ...800...51B}
{Boyer}, M.~L., {McQuinn}, K. B.~W., {Barmby}, P., {et~al.} 2015, \apj, 800,
  51, \dodoi{10.1088/0004-637X/800/1/51}

\bibitem[{{Boyer} {et~al.}(2017){Boyer}, {McQuinn}, {Groenewegen}, {Zijlstra},
  {Whitelock}, {van Loon}, {Sonneborn}, {Sloan}, {Skillman}, {Meixner},
  {McDonald}, {Jones}, {Javadi}, {Gehrz}, {Britavskiy}, \&
  {Bonanos}}]{2017ApJ...851..152B}
{Boyer}, M.~L., {McQuinn}, K.~B.~W., {Groenewegen}, M.~A.~T., {et~al.} 2017,
  \apj, 851, 152, \dodoi{10.3847/1538-4357/aa9892}

\bibitem[{{Cesarsky} {et~al.}(1978){Cesarsky}, {Laustsen}, {Lequeux},
  {Schuster}, \& {West}}]{1978A&A....65..153C}
{Cesarsky}, D.~A., {Laustsen}, S., {Lequeux}, J., {Schuster}, H.~E., \& {West},
  R.~M. 1978, \aap, 65, 153

\bibitem[{{Chambers} {et~al.}(2016){Chambers}, {Magnier}, {Metcalfe},
  {Flewelling}, {Huber}, {Waters}, {Denneau}, {Draper}, {Farrow}, {Finkbeiner},
  {Holmberg}, {Koppenhoefer}, {Price}, {Rest}, {Saglia}, {Schlafly}, {Smartt},
  {Sweeney}, {Wainscoat}, {Burgett}, {Chastel}, {Grav}, {Heasley}, {Hodapp},
  {Jedicke}, {Kaiser}, {Kudritzki}, {Luppino}, {Lupton}, {Monet}, {Morgan},
  {Onaka}, {Shiao}, {Stubbs}, {Tonry}, {White}, {Ba{\~n}ados}, {Bell},
  {Bender}, {Bernard}, {Boegner}, {Boffi}, {Botticella}, {Calamida},
  {Casertano}, {Chen}, {Chen}, {Cole}, {Deacon}, {Frenk}, {Fitzsimmons},
  {Gezari}, {Gibbs}, {Goessl}, {Goggia}, {Gourgue}, {Goldman}, {Grant},
  {Grebel}, {Hambly}, {Hasinger}, {Heavens}, {Heckman}, {Henderson}, {Henning},
  {Holman}, {Hopp}, {Ip}, {Isani}, {Jackson}, {Keyes}, {Koekemoer}, {Kotak},
  {Le}, {Liska}, {Long}, {Lucey}, {Liu}, {Martin}, {Masci}, {McLean}, {Mindel},
  {Misra}, {Morganson}, {Murphy}, {Obaika}, {Narayan}, {Nieto-Santisteban},
  {Norberg}, {Peacock}, {Pier}, {Postman}, {Primak}, {Rae}, {Rai}, {Riess},
  {Riffeser}, {Rix}, {R{\"o}ser}, {Russel}, {Rutz}, {Schilbach}, {Schultz},
  {Scolnic}, {Strolger}, {Szalay}, {Seitz}, {Small}, {Smith}, {Soderblom},
  {Taylor}, {Thomson}, {Taylor}, {Thakar}, {Thiel}, {Thilker}, {Unger},
  {Urata}, {Valenti}, {Wagner}, {Walder}, {Walter}, {Watters}, {Werner},
  {Wood-Vasey}, \& {Wyse}}]{2016arXiv161205560C}
{Chambers}, K.~C., {Magnier}, E.~A., {Metcalfe}, N., {et~al.} 2016, arXiv
  e-prints, arXiv:1612.05560.
\newblock \doarXiv{1612.05560}

\bibitem[{{Cook}(1987)}]{1987PhDT.........6C}
{Cook}, K.~H. 1987, PhD thesis, Arizona Univ., Tucson.

\bibitem[{{Demers} \& {Battinelli}(2002)}]{2002AJ....123..238D}
{Demers}, S., \& {Battinelli}, P. 2002, \aj, 123, 238, \dodoi{10.1086/324735}

\bibitem[{{Erben} {et~al.}(2005){Erben}, {Schirmer}, {Dietrich}, {Cordes},
  {Haberzettl}, {Hetterscheidt}, {Hildebrandt}, {Schmithuesen}, {Schneider},
  {Simon}, {Deul}, {Hook}, {Kaiser}, {Radovich}, {Benoist}, {Nonino}, {Olsen},
  {Prandoni}, {Wichmann}, {Zaggia}, {Bomans}, {Dettmar}, \&
  {Miralles}}]{2005AN....326..432E}
{Erben}, T., {Schirmer}, M., {Dietrich}, J.~P., {et~al.} 2005, Astronomische
  Nachrichten, 326, 432, \dodoi{10.1002/asna.200510396}

\bibitem[{{Gaia Collaboration} {et~al.}(2021){Gaia Collaboration}, {Brown},
  {Vallenari}, {Prusti}, {de Bruijne}, {Babusiaux}, {Biermann}, {Creevey},
  {Evans}, {Eyer}, {Hutton}, {Jansen}, {Jordi}, {Klioner}, {Lammers},
  {Lindegren}, {Luri}, {Mignard}, {Panem}, {Pourbaix}, {Randich}, {Sartoretti},
  {Soubiran}, {Walton}, {Arenou}, {Bailer-Jones}, {Bastian}, {Cropper},
  {Drimmel}, {Katz}, {Lattanzi}, {van Leeuwen}, {Bakker}, {Cacciari},
  {Casta{\~n}eda}, {De Angeli}, {Ducourant}, {Fabricius}, {Fouesneau},
  {Fr{\'e}mat}, {Guerra}, {Guerrier}, {Guiraud}, {Jean-Antoine Piccolo},
  {Masana}, {Messineo}, {Mowlavi}, {Nicolas}, {Nienartowicz}, {Pailler},
  {Panuzzo}, {Riclet}, {Roux}, {Seabroke}, {Sordo}, {Tanga}, {Th{\'e}venin},
  {Gracia-Abril}, {Portell}, {Teyssier}, {Altmann}, {Andrae}, {Bellas-Velidis},
  {Benson}, {Berthier}, {Blomme}, {Brugaletta}, {Burgess}, {Busso}, {Carry},
  {Cellino}, {Cheek}, {Clementini}, {Damerdji}, {Davidson}, {Delchambre},
  {Dell'Oro}, {Fern{\'a}ndez-Hern{\'a}ndez}, {Galluccio}, {Garc{\'\i}a-Lario},
  {Garcia-Reinaldos}, {Gonz{\'a}lez-N{\'u}{\~n}ez}, {Gosset}, {Haigron},
  {Halbwachs}, {Hambly}, {Harrison}, {Hatzidimitriou}, {Heiter},
  {Hern{\'a}ndez}, {Hestroffer}, {Hodgkin}, {Holl}, {Jan{\ss}en}, {Jevardat de
  Fombelle}, {Jordan}, {Krone-Martins}, {Lanzafame}, {L{\"o}ffler}, {Lorca},
  {Manteiga}, {Marchal}, {Marrese}, {Moitinho}, {Mora}, {Muinonen}, {Osborne},
  {Pancino}, {Pauwels}, {Petit}, {Recio-Blanco}, {Richards}, {Riello},
  {Rimoldini}, {Robin}, {Roegiers}, {Rybizki}, {Sarro}, {Siopis}, {Smith},
  {Sozzetti}, {Ulla}, {Utrilla}, {van Leeuwen}, {van Reeven}, {Abbas}, {Abreu
  Aramburu}, {Accart}, {Aerts}, {Aguado}, {Ajaj}, {Altavilla}, {{\'A}lvarez},
  {{\'A}lvarez Cid-Fuentes}, {Alves}, {Anderson}, {Anglada Varela}, {Antoja},
  {Audard}, {Baines}, {Baker}, {Balaguer-N{\'u}{\~n}ez}, {Balbinot}, {Balog},
  {Barache}, {Barbato}, {Barros}, {Barstow}, {Bartolom{\'e}}, {Bassilana},
  {Bauchet}, {Baudesson-Stella}, {Becciani}, {Bellazzini}, {Bernet}, {Bertone},
  {Bianchi}, {Blanco-Cuaresma}, {Boch}, {Bombrun}, {Bossini}, {Bouquillon},
  {Bragaglia}, {Bramante}, {Breedt}, {Bressan}, {Brouillet}, {Bucciarelli},
  {Burlacu}, {Busonero}, {Butkevich}, {Buzzi}, {Caffau}, {Cancelliere},
  {C{\'a}novas}, {Cantat-Gaudin}, {Carballo}, {Carlucci}, {Carnerero},
  {Carrasco}, {Casamiquela}, {Castellani}, {Castro-Ginard}, {Castro Sampol},
  {Chaoul}, {Charlot}, {Chemin}, {Chiavassa}, {Cioni}, {Comoretto}, {Cooper},
  {Cornez}, {Cowell}, {Crifo}, {Crosta}, {Crowley}, {Dafonte}, {Dapergolas},
  {David}, {David}, {de Laverny}, {De Luise}, {De March}, {De Ridder}, {de
  Souza}, {de Teodoro}, {de Torres}, {del Peloso}, {del Pozo}, {Delbo},
  {Delgado}, {Delgado}, {Delisle}, {Di Matteo}, {Diakite}, {Diener},
  {Distefano}, {Dolding}, {Eappachen}, {Edvardsson}, {Enke}, {Esquej}, {Fabre},
  {Fabrizio}, {Faigler}, {Fedorets}, {Fernique}, {Fienga}, {Figueras},
  {Fouron}, {Fragkoudi}, {Fraile}, {Franke}, {Gai}, {Garabato},
  {Garcia-Gutierrez}, {Garc{\'\i}a-Torres}, {Garofalo}, {Gavras}, {Gerlach},
  {Geyer}, {Giacobbe}, {Gilmore}, {Girona}, {Giuffrida}, {Gomel}, {Gomez},
  {Gonzalez-Santamaria}, {Gonz{\'a}lez-Vidal}, {Granvik},
  {Guti{\'e}rrez-S{\'a}nchez}, {Guy}, {Hauser}, {Haywood}, {Helmi}, {Hidalgo},
  {Hilger}, {H{\l}adczuk}, {Hobbs}, {Holland}, {Huckle}, {Jasniewicz},
  {Jonker}, {Juaristi Campillo}, {Julbe}, {Karbevska}, {Kervella}, {Khanna},
  {Kochoska}, {Kontizas}, {Kordopatis}, {Korn}, {Kostrzewa-Rutkowska},
  {Kruszy{\'n}ska}, {Lambert}, {Lanza}, {Lasne}, {Le Campion}, {Le Fustec},
  {Lebreton}, {Lebzelter}, {Leccia}, {Leclerc}, {Lecoeur-Taibi}, {Liao},
  {Licata}, {Lindstr{\o}m}, {Lister}, {Livanou}, {Lobel}, {Madrero Pardo},
  {Managau}, {Mann}, {Marchant}, {Marconi}, {Marcos Santos}, {Marinoni},
  {Marocco}, {Marshall}, {Martin Polo}, {Mart{\'\i}n-Fleitas}, {Masip},
  {Massari}, {Mastrobuono-Battisti}, {Mazeh}, {McMillan}, {Messina},
  {Michalik}, {Millar}, {Mints}, {Molina}, {Molinaro}, {Moln{\'a}r},
  {Montegriffo}, {Mor}, {Morbidelli}, {Morel}, {Morris}, {Mulone}, {Munoz},
  {Muraveva}, {Murphy}, {Musella}, {Noval}, {Ord{\'e}novic}, {Orr{\`u}},
  {Osinde}, {Pagani}, {Pagano}, {Palaversa}, {Palicio}, {Panahi}, {Pawlak},
  {Pe{\~n}alosa Esteller}, {Penttil{\"a}}, {Piersimoni}, {Pineau}, {Plachy},
  {Plum}, {Poggio}, {Poretti}, {Poujoulet}, {Pr{\v{s}}a}, {Pulone}, {Racero},
  {Ragaini}, {Rainer}, {Raiteri}, {Rambaux}, {Ramos}, {Ramos-Lerate}, {Re
  Fiorentin}, {Regibo}, {Reyl{\'e}}, {Ripepi}, {Riva}, {Rixon}, {Robichon},
  {Robin}, {Roelens}, {Rohrbasser}, {Romero-G{\'o}mez}, {Rowell}, {Royer},
  {Rybicki}, {Sadowski}, {Sagrist{\`a} Sell{\'e}s}, {Sahlmann}, {Salgado},
  {Salguero}, {Samaras}, {Sanchez Gimenez}, {Sanna}, {Santove{\~n}a},
  {Sarasso}, {Schultheis}, {Sciacca}, {Segol}, {Segovia}, {S{\'e}gransan},
  {Semeux}, {Shahaf}, {Siddiqui}, {Siebert}, {Siltala}, {Slezak}, {Smart},
  {Solano}, {Solitro}, {Souami}, {Souchay}, {Spagna}, {Spoto}, {Steele},
  {Steidelm{\"u}ller}, {Stephenson}, {S{\"u}veges}, {Szabados}, {Szegedi-Elek},
  {Taris}, {Tauran}, {Taylor}, {Teixeira}, {Thuillot}, {Tonello}, {Torra},
  {Torra}, {Turon}, {Unger}, {Vaillant}, {van Dillen}, {Vanel}, {Vecchiato},
  {Viala}, {Vicente}, {Voutsinas}, {Weiler}, {Wevers}, {Wyrzykowski}, {Yoldas},
  {Yvard}, {Zhao}, {Zorec}, {Zucker}, {Zurbach}, \&
  {Zwitter}}]{2021A&A...649A...1G}
{Gaia Collaboration}, {Brown}, A.~G.~A., {Vallenari}, A., {et~al.} 2021, \aap,
  649, A1, \dodoi{10.1051/0004-6361/202039657}

\bibitem[{{Girardi} {et~al.}(2005){Girardi}, {Groenewegen}, {Hatziminaoglou},
  \& {da Costa}}]{2005A&A...436..895G}
{Girardi}, L., {Groenewegen}, M.~A.~T., {Hatziminaoglou}, E., \& {da Costa}, L.
  2005, \aap, 436, 895, \dodoi{10.1051/0004-6361:20042352}

\bibitem[{{Goldman} {et~al.}(2019){Goldman}, {Boyer}, {McQuinn}, {Whitelock},
  {McDonald}, {van Loon}, {Skillman}, {Gehrz}, {Javadi}, {Sloan}, {Jones},
  {Groenewegen}, \& {Menzies}}]{2019ApJ...877...49G}
{Goldman}, S.~R., {Boyer}, M.~L., {McQuinn}, K.~B.~W., {et~al.} 2019, \apj,
  877, 49, \dodoi{10.3847/1538-4357/ab0965}

\bibitem[{{Gullieuszik} {et~al.}(2007){Gullieuszik}, {Rejkuba}, {Cioni},
  {Habing}, \& {Held}}]{2007A&A...475..467G}
{Gullieuszik}, M., {Rejkuba}, M., {Cioni}, M.~R., {Habing}, H.~J., \& {Held},
  E.~V. 2007, \aap, 475, 467, \dodoi{10.1051/0004-6361:20066848}

\bibitem[{{Hamedani Golshan} {et~al.}(2017){Hamedani Golshan}, {Javadi}, {van
  Loon}, {Khosroshahi}, \& {Saremi}}]{2017MNRAS.466.1764H}
{Hamedani Golshan}, R., {Javadi}, A., {van Loon}, J.~T., {Khosroshahi}, H., \&
  {Saremi}, E. 2017, \mnras, 466, 1764, \dodoi{10.1093/mnras/stw3174}

\bibitem[{Hashemi {et~al.}(2018)Hashemi, Javadi, \&
  van Loon}]{10.1093/mnras/sty3450}
Hashemi, S.~A., Javadi, A., \& van Loon, J.~T. 2018, \mnras, 483, 4751,
  \dodoi{10.1093/mnras/sty3450}

\bibitem[{{Haynes}(2019)}]{2019IAUS..344....3H}
{Haynes}, M.~P. 2019, in Dwarf Galaxies: From the Deep Universe to the Present,
  ed. K.~B.~W. {McQuinn} \& S.~{Stierwalt}, Vol. 344, 3--16,
  \dodoi{10.1017/S1743921319000073}

\bibitem[{{Higgs} {et~al.}(2016){Higgs}, {McConnachie}, {Irwin}, {Bate},
  {Lewis}, {Walker}, {C{\^o}t{\'e}}, {Venn}, \&
  {Battaglia}}]{2016MNRAS.458.1678H}
{Higgs}, C.~R., {McConnachie}, A.~W., {Irwin}, M., {et~al.} 2016, \mnras, 458,
  1678, \dodoi{10.1093/mnras/stw257}

\bibitem[{{Hunter} {et~al.}(2012){Hunter}, {Ficut-Vicas}, {Ashley}, {Brinks},
  {Cigan}, {Elmegreen}, {Heesen}, {Herrmann}, {Johnson}, {Oh}, {Rupen},
  {Schruba}, {Simpson}, {Walter}, {Westpfahl}, {Young}, \&
  {Zhang}}]{2012AJ....144..134H}
{Hunter}, D.~A., {Ficut-Vicas}, D., {Ashley}, T., {et~al.} 2012, \aj, 144, 134,
  \dodoi{10.1088/0004-6256/144/5/134}

\bibitem[{{Javadi} {et~al.}(2017){Javadi}, {van Loon}, {Khosroshahi},
  {Tabatabaei}, {Hamedani Golshan}, \& {Rashidi}}]{2017MNRAS.464.2103J}
{Javadi}, A., {van Loon}, J.~T., {Khosroshahi}, H.~G., {et~al.} 2017, \mnras,
  464, 2103, \dodoi{10.1093/mnras/stw2463}

\bibitem[{{Javadi} {et~al.}(2011{\natexlab{a}}){Javadi}, {van Loon}, \&
  {Mirtorabi}}]{2011MNRAS.411..263J}
{Javadi}, A., {van Loon}, J.~T., \& {Mirtorabi}, M.~T. 2011{\natexlab{a}},
  \mnras, 411, 263, \dodoi{10.1111/j.1365-2966.2010.17678.x}

\bibitem[{{Javadi} {et~al.}(2011{\natexlab{b}}){Javadi}, {van Loon}, \&
  {Mirtorabi}}]{8214023}
---. 2011{\natexlab{b}}, \mnras, 414, 3394

\bibitem[{{Karachentsev} {et~al.}(1999){Karachentsev}, {Aparicio}, \&
  {Makarova}}]{1999A&A...352..363K}
{Karachentsev}, I., {Aparicio}, A., \& {Makarova}, L. 1999, \aap, 352, 363.
\newblock \doarXiv{astro-ph/9910136}

\bibitem[{{Kirby} {et~al.}(2017){Kirby}, {Rizzi}, {Held}, {Cohen}, {Cole},
  {Manning}, {Skillman}, \& {Weisz}}]{2017ApJ...834....9K}
{Kirby}, E.~N., {Rizzi}, L., {Held}, E.~V., {et~al.} 2017, \apj, 834, 9,
  \dodoi{10.3847/1538-4357/834/1/9}

\bibitem[{{Kroupa}(2001)}]{2001MNRAS.322..231K}
{Kroupa}, P. 2001, \mnras, 322, 231, \dodoi{10.1046/j.1365-8711.2001.04022.x}

\bibitem[{{Landolt}(1992)}]{1992AJ....104..340L}
{Landolt}, A.~U. 1992, \aj, 104, 340, \dodoi{10.1086/116242}

\bibitem[{{Lee} {et~al.}(1993){Lee}, {Freedman}, \&
  {Madore}}]{1993ApJ...417..553L}
{Lee}, M.~G., {Freedman}, W.~L., \& {Madore}, B.~F. 1993, \apj, 417, 553,
  \dodoi{10.1086/173334}

\bibitem[{{Longmore} {et~al.}(1978){Longmore}, {Hawarden}, {Webster}, {Goss},
  \& {Mebold}}]{1978MNRAS.183P..97L}
{Longmore}, A.~J., {Hawarden}, T.~G., {Webster}, B.~L., {Goss}, W.~M., \&
  {Mebold}, U. 1978, \mnras, 183, 97P, \dodoi{10.1093/mnras/183.1.97P}

\bibitem[{{Marigo} {et~al.}(2017){Marigo}, {Girardi}, {Bressan}, {Rosenfield},
  {Aringer}, {Chen}, {Dussin}, {Nanni}, {Pastorelli}, {Rodrigues}, {Trabucchi},
  {Bladh}, {Dalcanton}, {Groenewegen}, {Montalb{\'a}n}, \&
  {Wood}}]{2017ApJ...835...77M}
{Marigo}, P., {Girardi}, L., {Bressan}, A., {et~al.} 2017, \apj, 835, 77,
  \dodoi{10.3847/1538-4357/835/1/77}

\bibitem[{{McConnachie}(2012)}]{2012AJ....144....4M}
{McConnachie}, A.~W. 2012, \aj, 144, 4, \dodoi{10.1088/0004-6256/144/1/4}

\bibitem[{{McQuinn} {et~al.}(2017){McQuinn}, {Boyer}, {Mitchell}, {Skillman},
  {Gehrz}, {Groenewegen}, {McDonald}, {Sloan}, {van Loon}, {Whitelock}, \&
  {Zijlstra}}]{2017ApJ...834...78M}
{McQuinn}, K. B.~W., {Boyer}, M.~L., {Mitchell}, M.~B., {et~al.} 2017, \apj,
  834, 78, \dodoi{10.3847/1538-4357/834/1/78}

\bibitem[{{Momany} {et~al.}(2002){Momany}, {Held}, {Saviane}, \&
  {Rizzi}}]{2002A&A...384..393M}
{Momany}, Y., {Held}, E.~V., {Saviane}, I., \& {Rizzi}, L. 2002, \aap, 384,
  393, \dodoi{10.1051/0004-6361:20020047}

\bibitem[{{Momany} {et~al.}(2005){Momany}, {Held}, {Saviane}, {Bedin},
  {Gullieuszik}, {Clemens}, {Rizzi}, {Rich}, \&
  {Kuijken}}]{2005A&A...439..111M}
{Momany}, Y., {Held}, E.~V., {Saviane}, I., {et~al.} 2005, \aap, 439, 111,
  \dodoi{10.1051/0004-6361:20052747}

\bibitem[{{Momany} {et~al.}(2014){Momany}, {Clemens}, {Bedin}, {Gullieuszik},
  {Held}, {Saviane}, {Zaggia}, {Monaco}, {Montalto}, {Rich}, \&
  {Rizzi}}]{2014A&A...572A..42M}
{Momany}, Y., {Clemens}, M., {Bedin}, L.~R., {et~al.} 2014, \aap, 572, A42,
  \dodoi{10.1051/0004-6361/201424055}

\bibitem[{Navabi {et~al.}(2021)Navabi, Saremi, Javadi, Noori, van Loon,
  Khosroshahi, McDonald, Alizadeh, Danesh, Gozaliasl, Molaeinezhad, Parto, \&
  Raouf}]{Navabi_2021}
Navabi, M., Saremi, E., Javadi, A., {et~al.} 2021, The Astrophysical Journal,
  910, 127, \dodoi{10.3847/1538-4357/abdec1}

\bibitem[{{Parto} {et~al.}(2021){Parto}, {Dehghani}, {Javadi}, {Saremi}, {van
  Loon}, {Khosroshahi}, {Taghi Mirtorabi}, {Abdollahi}, {Gholami}, {Azim
  Hashemi}, {Navabi}, {Noori}, {Taefi Aghdam}, {Torki}, \&
  {Vafaeizade}}]{2021arXiv210110874P}
{Parto}, T., {Dehghani}, S., {Javadi}, A., {et~al.} 2021, arXiv e-prints,
  arXiv:2101.10874.
\newblock \doarXiv{2101.10874}

\bibitem[{{Rezaeikh} {et~al.}(2014){Rezaeikh}, {Javadi}, {Khosroshahi}, \& {van
  Loon}}]{2014MNRAS.445.2214R}
{Rezaeikh}, S., {Javadi}, A., {Khosroshahi}, H., \& {van Loon}, J.~T. 2014,
  \mnras, 445, 2214, \dodoi{10.1093/mnras/stu1807}

\bibitem[{{Sakai} {et~al.}(1996){Sakai}, {Madore}, \&
  {Freedman}}]{1996ApJ...461..713S}
{Sakai}, S., {Madore}, B.~F., \& {Freedman}, W.~L. 1996, \apj, 461, 713,
  \dodoi{10.1086/177096}

\bibitem[{Saremi {et~al.}(2021)Saremi, Javadi, Navabi, van Loon, Khosroshahi,
  Arbab, \& McDonald}]{Saremi_2021}
Saremi, E., Javadi, A., Navabi, M., {et~al.} 2021, The Astrophysical Journal,
  923, 164, \dodoi{10.3847/1538-4357/ac2d96}

\bibitem[{{Saremi} {et~al.}(2017){Saremi}, {Javadi}, {van Loon}, {Khosroshahi},
  {Abedi}, {Bamber}, {Hashemi}, {Nikzat}, \& {Molaei
  Nezhad}}]{2017JPhCS.869a2068S}
{Saremi}, E., {Javadi}, A., {van Loon}, J.~T., {et~al.} 2017, in Journal of
  Physics Conference Series, Vol. 869, Journal of Physics Conference Series,
  012068, \dodoi{10.1088/1742-6596/869/1/012068}

\bibitem[{{Saremi} {et~al.}(2020){Saremi}, {Javadi}, {Th. van Loon},
  {Khosroshahi}, {Molaeinezhad}, {McDonald}, {Raouf}, {Danesh}, {Bamber},
  {Short}, {Su{\'a}rez-Andr{\'e}s}, {Clavero}, \&
  {Gozaliasl}}]{2020ApJ...894..135S}
{Saremi}, E., {Javadi}, A., {Th. van Loon}, J., {et~al.} 2020, \apj, 894, 135,
  \dodoi{10.3847/1538-4357/ab88a2}

\bibitem[{{Saviane} {et~al.}(2002){Saviane}, {Rizzi}, {Held}, {Bresolin}, \&
  {Momany}}]{2002A&A...390...59S}
{Saviane}, I., {Rizzi}, L., {Held}, E.~V., {Bresolin}, F., \& {Momany}, Y.
  2002, \aap, 390, 59, \dodoi{10.1051/0004-6361:20020750}

\bibitem[{{Schlegel} {et~al.}(1998){Schlegel}, {Finkbeiner}, \&
  {Davis}}]{1998ApJ...500..525S}
{Schlegel}, D.~J., {Finkbeiner}, D.~P., \& {Davis}, M. 1998, \apj, 500, 525,
  \dodoi{10.1086/305772}

\bibitem[{{Simon}(2019)}]{2019ARA&A..57..375S}
{Simon}, J.~D. 2019, \araa, 57, 375,
  \dodoi{10.1146/annurev-astro-091918-104453}

\bibitem[{{Stetson}(1987)}]{1987PASP...99..191S}
{Stetson}, P.~B. 1987, \pasp, 99, 191, \dodoi{10.1086/131977}

\bibitem[{{Stetson}(1990)}]{1990PASP..102..932S}
---. 1990, \pasp, 102, 932, \dodoi{10.1086/132719}

\bibitem[{{Stetson}(1993)}]{1993spct.conf..291S}
{Stetson}, P.~B. 1993, in IAU Colloq. 136: Stellar Photometry - Current
  Techniques and Future Developments, ed. C.~J. {Butler} \& I.~{Elliott}, 291

\bibitem[{{Stetson}(1994)}]{1994PASP..106..250S}
---. 1994, \pasp, 106, 250, \dodoi{10.1086/133378}

\bibitem[{{Stetson}(1996)}]{1996PASP..108..851S}
---. 1996, \pasp, 108, 851, \dodoi{10.1086/133808}

\bibitem[{{Strobel} {et~al.}(1991){Strobel}, {Hodge}, \&
  {Kennicutt}}]{1991ApJ...383..148S}
{Strobel}, N.~V., {Hodge}, P., \& {Kennicutt}, Robert~C., J. 1991, \apj, 383,
  148, \dodoi{10.1086/170771}

\bibitem[{{Weisz} {et~al.}(2014){Weisz}, {Dolphin}, {Skillman}, {Holtzman},
  {Gilbert}, {Dalcanton}, \& {Williams}}]{2014ApJ...789..147W}
{Weisz}, D.~R., {Dolphin}, A.~E., {Skillman}, E.~D., {et~al.} 2014, \apj, 789,
  147, \dodoi{10.1088/0004-637X/789/2/147}

\bibitem[{{Whitelock} {et~al.}(2018){Whitelock}, {Menzies}, {Feast}, \&
  {Marigo}}]{2018MNRAS.473..173W}
{Whitelock}, P.~A., {Menzies}, J.~W., {Feast}, M.~W., \& {Marigo}, P. 2018,
  \mnras, 473, 173, \dodoi{10.1093/mnras/stx2275}

\bibitem[{{Young} \& {Lo}(1997)}]{1997ApJ...490..710Y}
{Young}, L.~M., \& {Lo}, K.~Y. 1997, \apj, 490, 710, \dodoi{10.1086/304909}

\end{thebibliography}

\bibliographystyle{aasjournal}

\begin{table}[]
	\centering
	\appendix{\label{app}}
	\caption{\label{table:starsparam} Photometric properties of the detected LPV candidates in SagDIG.} 
	\begin{tabular}{ccccccccccccc}
		\hline
		ID   & R.A. (J2000) & Dec. (J2000) & V      & $err_{V}$ & i      & $err_{i}$ & n$^a$ & m$^b$ & J index & K index & L index & Amplitude \\ 
		\hline
		5129 & 19:30:04.424 & -17:40:04.54 & 22.744 & 0.1420    & 21.718 & 0.0668    & 3 & 7 & 1.111   & 0.918   & 1.306   & 0.588     \\
		5131 & 19:30:04.407 & -17:41:12.93 & 22.597 & 0.2485    & 21.409 & 0.1556    & 3 & 7 & 0.994   & 0.757   & 1.297   & 1.437     \\
		5408 & 19:30:02.520 & -17:41:17.53 & 21.757 & 0.1950    & 21.073 & 0.0975    & 4 & 6 & 1.413   & 0.660   & 1.347   & 0.584     \\
		5415 & 19:30:02.460 & -17:40:21.59 & 22.460 & 0.2600    & 21.600 & 0.0866    & 3 & 7 & 1.627   & 0.715   & 1.422   & 0.674     \\
		5420 & 19:30:02.453 & -17:41:15.33 & 22.074 & 0.2926    & 21.310 & 0.1117    & 4 & 7 & 1.536   & 0.597   & 1.676   & 0.784     \\
		5444 & 19:30:02.278 & -17:41:22.98 & 21.674 & 0.2052    & 21.000 & 0.1062    & 1 & 7 & 1.796   & 0.660   & 1.717   & 1.211     \\
		5514 & 19:30:02.026 & -17:40:19.41 & 19.722 & 0.0183    & 19.416 & 0.0150    & 4 & 7 & 2.078   & 0.821   & 2.005   & 0.157     \\
		5560 & 19:30:01.683 & -17:40:04.14 & 21.972 & 0.1131    & 20.611 & 0.1020    & 4 & 7 & 3.153   & 0.616   & 4.197   & 0.596     \\
		5562 & 19:30:01.675 & -17:41:06.57 & 22.666 & 0.2053    & 21.151 & 0.1093    & 3 & 7 & 2.202   & 0.798   & 2.270   & 0.719     \\
		5578 & 19:30:01.606 & -17:40:35.25 & 23.169 & 0.2034    & 21.344 & 0.0666    & 1 & 7 & 0.901   & 0.806   & 1.257   & 0.424     \\
		5580 & 19:30:01.594 & -17:40:57.83 & 22.204 & 0.1967    & 21.860 & 0.1718    & 4 & 7 & 0.468   & 0.627   & 1.188   & 1.195     \\
		5699 & 19:30:01.015 & -17:40:52.63 & 22.394 & 0.1176    & 20.629 & 0.0390    & 4 & 7 & 1.247   & 0.679   & 1.213   & 0.313     \\
		5749 & 19:30:00.687 & -17:41:00.56 & 99.999 & 9.9999    & 22.038 & 0.2991    & 0 & 6 & 3.593   & 0.699   & 2.942   & 1.801     \\
		5808 & 19:30:00.241 & -17:41:12.25 & 21.499 & 0.0783    & 21.050 & 0.0707    & 4 & 7 & 1.434   & 0.680   & 1.503   & 0.558     \\
		5831 & 19:30:00.072 & -17:41:35.17 & 23.003 & 0.1206    & 21.021 & 0.0946    & 2 & 7 & 1.350   & 0.602   & 1.322   & 0.574     \\
		5865 & 19:29:59.766 & -17:40:34.08 & 22.333 & 0.1022    & 20.555 & 0.0352    & 4 & 7 & 1.273   & 0.758   & 1.256   & 0.275     \\
		5875 & 19:29:59.708 & -17:39:54.65 & 21.272 & 0.1519    & 20.812 & 0.1451    & 1 & 6 & 2.176   & 0.599   & 2.162   & 0.991     \\
		5909 & 19:29:59.442 & -17:39:59.22 & 22.345 & 0.3693    & 21.886 & 0.1294    & 3 & 7 & 1.383   & 0.833   & 1.631   & 0.927     \\
		6092 & 19:29:58.505 & -17:40:42.12 & 21.966 & 0.1130    & 21.468 & 0.0712    & 4 & 7 & 1.437   & 0.828   & 1.431   & 0.538     \\
		6161 & 19:29:57.944 & -17:40:17.46 & 21.311 & 0.1018    & 20.597 & 0.1429    & 4 & 7 & 0.089   & 0.514   & 1.313   & 0.927     \\
		6217 & 19:29:57.553 & -17:40:20.30 & 22.042 & 0.2068    & 21.505 & 0.1372    & 4 & 7 & 1.205   & 0.617   & 2.067   & 0.790     \\
		6941 & 19:29:52.909 & -17:40:32.73 & 23.020 & 0.2469    & 21.103 & 0.0769    & 3 & 7 & 1.922   & 0.720   & 2.285   & 0.497     \\
		6345 & 19:29:56.801 & -17:40:21.69 & 22.635 & 0.1134    & 22.057 & 0.1027    & 4 & 7 & 1.133   & 0.694   & 1.135   & 0.689     \\
		4727 & 19:30:07.209 & -17:40:27.10 & 23.399 & 0.3766    & 20.910 & 0.0523    & 2 & 7 & 1.144   & 0.806   & 1.223   & 0.351     \\
		4822 & 19:30:06.648 & -17:40:09.73 & 21.084 & 0.1367    & 20.556 & 0.0933    & 4 & 7 & 1.789   & 0.662   & 1.542   & 1.158     \\
		6578 & 19:29:55.074 & -17:41:14.11 & 19.438 & 0.3179    & 18.757 & 0.2712    & 4 & 6 & 1.088   & 0.844   & 1.471   & 1.541     \\
		5900 & 19:29:59.500 & -17:39:50.48 & 20.425 & 0.0462    & 19.896 & 0.1353    & 4 & 7 & 1.067   & 0.658   & 1.236   & 0.744     \\ \hline
	\end{tabular}
	
    \raggedright
	$^a$ n is the number of observations in the $V$-band.\\
	$^b$ m is the number of observations in the $i$-band.
\end{table}

\end{document}